\documentclass[graybox, nosecnum]{svmult}

\usepackage{mathptmx}       
\usepackage{helvet}         
\usepackage{courier}        
\usepackage{type1cm}        
\usepackage{makeidx}         
\usepackage{graphicx}        
\usepackage{multicol}        
\usepackage[bottom]{footmisc}
\usepackage{hyperref}        
\usepackage{soul}            
\usepackage{amsmath,amssymb}
\usepackage[english]{babel}
\usepackage{blindtext}
\usepackage{rotating}
\usepackage{amsfonts}
\usepackage{booktabs}
\usepackage{siunitx}
\usepackage{wasysym}
\usepackage{adjustbox}
\usepackage{MnSymbol}
\usepackage{enumitem}
\usepackage{threeparttable}
\hypersetup{colorlinks=true,urlcolor=blue}
\usepackage[square,numbers]{natbib}
\makeindex             

\begin{document}
\title*{Future developments in ground-based gamma-ray astronomy}
\author{Ulisses Barres de Almeida \thanks{corresponding author} and Martin Tluczykont}
\institute{Ulisses Barres de Almeida \at Brazilian Center for Physics Research (CBPF), Rua Dr. Xavier Sigaud 150, 22290-180 Rio de Janeiro, Brazil. \email{ulisses@cbpf.br}
\and Martin Tluczykont \at Institute of Experimental Physics, University of Hamburg, Luruper Chaussee 149, 22761 Hamburg, Germany. \email{martin.tluczykont@physik.uni-hamburg.de}}
%
%
\maketitle
\abstract{Ground-based $\gamma$-ray astronomy is a powerful tool to study cosmic-ray physics, providing a diagnostic of the high-energy processes at work in the most extreme astrophysical accelerators of the universe. Ground-based $\gamma$-ray detectors apply a number of experimental techniques to measure the products of air showers induced by the primary $\gamma$-rays over a wide energy range, from about 30 GeV to few PeV. These are based either on the measurement of the atmospheric Cherenkov light induced by the air showers, or the direct detection of the shower's secondary particles at ground level. Thanks to the recent development of new and highly sensitive ground-based $\gamma$-ray detectors, important scientific results are emerging which motivate new experimental proposals, at various stages of implementation. In this chapter we will present the current expectations for future experiments in the field.}

\vspace{0.5cm}

\newpage

\section{Introduction}
After several decades of incremental attempts, and following the rapid developments witnessed since the early years of this Century, the field of ground-based $\gamma$-ray astronomy has reached the status of 'real astronomy', consolidating a unique observational window into the sky which spans 5 decades in energy, from a few tens of GeV to the PeV, and several decades of typical flux difference between the low- and high-energy ends of the spectrum. The application of different types of techniques and technologies is required to probe this vast range of parameters, all based on the detection of the secondary products of the $\gamma$ ray-initiated extensive air showers (EAS). The main instrumental constraints are the requirement for large detection areas and the capability to efficiently suppress the much stronger background of cosmic ray (CR) protons.  

The two experimental approaches available are the air-Cherenkov technique, which observes the optical Cherenkov photons generated by shower particles traversing the atmosphere, and the particle detector arrays, which directly detect (sample) the secondary particles of the shower front at ground, and traditionally operate at higher energies than the air-Cherenkov instruments. The maturity achieved in the field results from the fact that the principal observatories of both classes have assembled a set of instrumental characteristics which proved critical in fulfilling the potentialities of ground-based observations. This means, in general, large array areas with sizes bigger than the shower footprint, and with a dense instrumental coverage, as well as powerful CR suppression factors, approaching the range of 10$^{-5}$, meaning 1 residual CR in 10,000, at least in the most performing ranges of operation of each technique. 

These achievements are the reason we can now speak of real astronomy at very- and ultra-high energies (VHE, 0.3-300 TeV, and UHE, above 300 TeV, respectively). They have enabled instruments to produce skymaps with resolutions superior to $\sim$ 0.1$^\circ$, to construct well-sampled source spectra spanning various decades in energy, up to several 100s of TeV, as well as explore flux variability with light-curves on sub-hour timescales, and as short as minutes for the brightest transients. Today, ground-based $\gamma$-ray astronomy is achieving the combined mark of nearly 300 detected sources.

The different approaches between air-Cherenkov instruments, in particular imaging air-Cherenkov telescopes (IACTs), and ground particle arrays, highlight complementarities that go beyond the typical observational energy ranges, and stress the importance of operating both types of instruments contemporaneously and in synergy. In this sense, to secure an adequate global latitude-longitude coverage with both experimental approaches is a main goal of the field for the near future. From the point of view of ground particle arrays, this means the installation of the first-ever instrument of its kind in the Southern Hemisphere, as is being targeted by a number of proposals from various groups. 

Improving background rejection, especially at the extremes of the energy range, is another fundamental challenge. In the (sub-)PeV domain, LHAASO recently demonstrated that near-background free operations (i.e., CR suppression factors close to $\sim 10^{-5}$) are required to effectively probe PeVatron accelerators~\cite{ZhenCao21PeV}. To achieve that in more cost-effective ways will depend on novel solutions for a km$^2$-scale array with large muon detection area, and is a major challenge for future Southern Hemisphere particle arrays.
Solutions using hybrid detector setups based on a combination of air-Cherenkov imaging and timing, and particle detection, such as realized by TAIGA, are also investigating a novel, cost-effective way to probe the VHE to UHE energy domain. These CR suppression factors at the highest energies are also pursued by IACTs and will, for example, benefit from the improved imaging and gamma-ray PSF of the Cherenkov Telescope Array (CTA), as well as the advanced analysis possibilities that are opened by such improvements.

The frontier towards the lowest energies -- that is, to reach below several tens of GeV for IACTs, and to approach the 100\,GeV threshold for ground particle arrays -- is another primary goal targeted by future proposals. The science motivation is focused on transient and extra-galactic sources, especially in the context of multi-messenger astrophysics and the follow-up of gravitational wave and neutrino event counterparts. For wide-field particle arrays, this will require building high fill-factor ($> 50\%$) instruments at even higher altitudes, approaching the 5 km a.s.l., and would enable to build effective VHE gamma-ray monitoring and trigger instruments that bridge the gap with satellite-based facilities. For IACTs, various technical developments ongoing in the context of CTA suggest that the O(10\,GeV) threshold is within reach. 

To achieve significant improvements in the angular resolution of ground-based gamma-ray observatories is an additional and essential goal, that would enhance science synergies not only between the VHE and UHE domains, but also with multi-band astronomical instrumentation. High fill-factor ground particle arrays with improved timing resolution are required for that, as well as large stereoscopic IACT arrays with highly-pixelated cameras and improved optics, as proposed for CTA. 

Finally, a key issue for the technological developments in all fronts is to keep production and deployment / maintenance costs as low as possible, so to render denser, larger-scale arrays viable.

\vspace{0.5cm}

\begin{figure}[t]
    \centering
    \includegraphics[width=\textwidth]{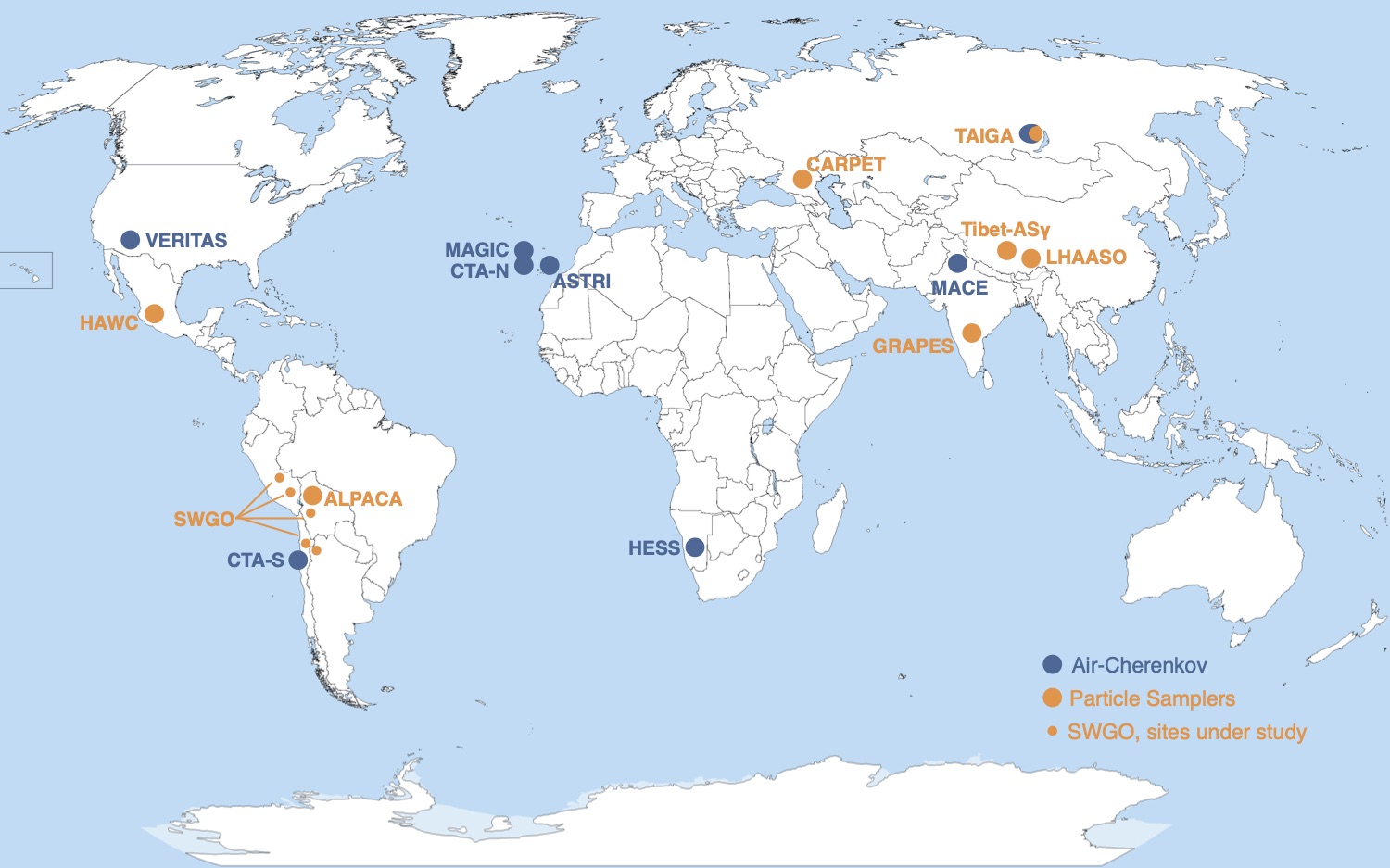}
    \caption{Geographic distribution of current and planned ground-based $\gamma$-ray facilities. In the case of SWGO, the potential site locations currently under investigation are indicated~\cite{DoroICRC21}.}
    \label{fig:FacilitiesMap}
\end{figure}

In this Chapter we will present current expectations for the future experiments in the field. For major endeavours that have started operations recently, or are at an advanced stage of development, such as LHAASO~\cite{ZhenCao21Nat} and CTA~\cite{Knodlseder20}, see chapters "Current Particle Detector Arrays in Gamma-ray Astronomy" and "The Cherenkov Telescope Array (CTA)", respectively. Likewise, ongoing upgrades of current experiments with $\gamma$-ray observation capabilities, such as the scintillator arrays CARPET-3~\cite{Kudzhaev21}, at the site of the Baksan Neutrino Observatory, in Russia, and GRAPES-3~\cite{MohantyICRC21}, located in Ooty, India, will not be discussed here, and the reader is again referred to the chapter "Current Particle Detector Arrays in Gamma-ray Astronomy"  for more details on these techniques. For completion, these experiments are nevertheless presented in Table~\ref{tab:Summary} and Figure~\ref{fig:FacilitiesMap}, which summarise the instrumental characteristics and geographic distribution of selected current and planned facilities.

After a brief overview of the experimental techniques, aimed at building a common background for the ensuing discussions, we will present in detail the efforts towards the Siberian Cosmic- and $\gamma$-ray detector, TAIGA, followed by the proposals for installation of a ground-based particle array detector in the Southern Hemisphere, and finally the proposals for future IACTs beyond CTA. The criteria driving our choice of which experimental proposals to present was their appearance in the 2019 and 2021 editions of the International Cosmic Ray Conference (ICRC). The text reflects the status of the field as of the end of
2022, when this article was compiled.

\subsection{Overview of Techniques}

{\bf Extensive Air Showers}

Cosmic rays and gamma rays initiate relativistic particle cascades in the Earth's atmosphere. The energy of the primary particle is transferred via high energy interactions (e.g. pair production, bremsstrahlung, etc.) to secondary cascade particles. These particles also create Cherenkov light which can be detected on the ground, either by measuring the density of the Cherenkov light distribution (wave-front sampling technique), or by imaging the Cherenkov light emitted from the cascade. The number of secondary particles grows, until a critical energy per particle is reached, below which ionization losses become dominant. Such particle cascades are called extended air showers (EAS), and are the tool provided by nature to ground-based gamma-ray astronomers.

One of the main tasks in experimental gamma-ray astronomy is the separation between gamma-rays and the dominant hadronic background. While the development of gamma-ray induced EAS can be modeled mainly based on the pair production and bremsstrahlung processes, the development of hadronic EAS is governed by the strong interaction. Due to the greater transverse momentum transferred to pions in strong interactions, hadronic EAS are wider (e.g. \cite{Hillas1996}) and can also have pronounced sub-structures, due to electromagnetic sub-showers initiated by gamma-rays from the decay of neutral pions. Furthermore, the shower maximum of air showers (correlated with the first interaction) depends on the nature of the EAS-initiating particle. In general, hadronic EAS develop their maximum deeper into the atmosphere than gamma-rays with similar energies. This leads to observable differences in the structure of the collected Cherenkov light or secondary particles on the ground, and to differences in the image shape between gamma-rays and hadrons.

With a sufficiently high energy of the primary particle, the EAS can reach the ground before the secondary particles lose all their energy by ionization. Particle detectors on the observation level, usually placed at high altitude, can then be used to measure the secondary EAS particle distributions and arrival times on the ground.
Another approach is to measure the Cherenkov light emitted by the secondary EAS particles, which is much less attenuated by the atmosphere and can reach down to lower altitudes. By measuring the air Chrenkov yield of the EAS, atmospheric Cherenkov detectors can record the arrival times of the shower front and measure the Cherenkov light lateral density profile. In the case of imaging telescopes, the most successful ground-based gamma-ray technique, the entire longitudinal profile of the EAS can be imaged.  
Finally, fluorescence or radio measurements are also used for EAS observations. 

In the following, we briefly introduce the main concepts behind the particle sampling and air Cherenkov techniques, both being the methods implemented by the experiments presented here. For a detailed description of detection principles in ground-based $\gamma$-ray astronomy, see chapter "How to Detect Gamma-rays from the Ground: An Introduction to the Detection Concepts".

\begin{sidewaystable}{b}
\begin{threeparttable}[b]
\caption{\rule{0pt}{9cm}\\Characteristics of proposed ground-based $\gamma$-ray facilities. Acronyms stand for Kilometer-square array (KM2A), Large sized telescope (LST), Mid sized telescope (MST), Muon detector (MD), Proportional counter (PRC), Small sized telescope (SST), Surface detector (SD), Underground detector (UD), Water Cherenkov detector (WCD), Wide-field Cherenkov Telescope  (WFCT).\\
\rule{35cm}{0pt}}
\small
\begin{tabular}{p{1.5cm}p{2.3cm}p{2cm}p{2.cm}p{3.cm}p{2.7cm}p{2.5cm}p{2.7cm}p{2.cm}p{0.7cm}} \toprule
Observatory & Category & Status & Location & Technology & Array Area & Unit Spacing & Photosensor & Muon Detection & Ref. \\ \midrule
ALPACA  & Particle Sampler         & 1/4 array (2021)                      & 4.7\,km a.s.l.           & Scintillator SD                       & SD array = 82,800\,m$^2$ & SD = 15\,m                            & (SD) fast-timing PMT     & WCD array                             & ~\cite{SakoICRC21} \\
&                          & 1/2 array (2022)              & Chacaltaya               & Water Cherenkov UD            & MD array = 5,400\,m$^2$  & 8$\times$896\,m$^2$ MD clus.  & (MD) 20" PMT             & Underground                   & \\\midrule
CARPET-3& Particle Sampler         & Upgrade (2022)                        & 1.7\,km a.s.l.           & Scintillator SD/UD                    & 200\,m$^2$               & Dense carpet                          & 6" PMT (FEU-49)          & Scintillator                          & ~\cite{Kudzhaev21} \\
&                          &                               & Baksan                   &                               & 600\,m$^2$               & Continuous                    & 6" PMT (FEU-49)          & Underground                   & \\\midrule
CoMET   & Particle Sampler         & R\&D Phase                            & 5.1\,km a.s.l.           & Water Cherenkov                       & 20,000\,m$^2$            & WCD $\approx$ 4\,m$^b$                & (WCD) 8" PMT             & Surface Array                         & ~\cite{MezekICRC21} \\
& Air Cherenkov            &                               & Andes$^a$                & Timing Array                  &                          & ACT $\approx$ 7\,m$^b$        & (ACT) 8$\times$3" PMT.   & Scintillator layer            & \\\midrule
GRAPES-3& Particle Sampler         & Upgrade                               & 2.2\,km a.s.l.           & Scintillator SD                       & 25,000\,m$^2$            & 8\,m                                  & 2" PMT                   & Proportional Ctr.                     & ~\cite{MohantyICRC21} \\
&                          & (Ongoing, MD)                 & Ooty                     & Proportional Ctr. UD          & 560\,m$^2$               & 16 $\times$ 35\,m$^2$ mod.    & 6\,m $\times$ 0.01$m^2$ tubes  & Underground             & \\\midrule
& Particle Sampler         &                               &                          & Water Cherenkov (SD)          & $\approx$ 80,000\,m$^2$  & 3$\times$WCDA pools           & 1.5"+8" PMT$^c$          &                               & \\
LHAASO  & Particle Sampler         & Completed                             & 4.4\,km a.s.l.           & Scintillator (SD)                     & $\approx$ 1\,km$^2$      & 15\,m                                 & 1.5" PMT.                & Underground                           & ~\cite{ZhenCao21Nat} \\
& Particle Sampler         & (2021)                        &(Mt. Haizi).              & Water Cherenkov (UD)          & $\approx$ 1\,km$^2$      & 30\,m                         & 8" PMT                   & (WCD array)                   &  \\
& Air Cherenkov            &                               &                          & Calorimetry                   & 256\,deg$^2$ FoV         & 18 WFCT                       & 32$times$32 1.2" SiPM    &                               & \\ \midrule
STACEX  & Particle Sampler         & R\&D Phase                            & $>$ 4\,km a.s.l.         & Resistive Plate Chamber               & $\approx$ 22,000\,m$^2$  & Dense carpet                          & 8" PMT                   & WCD array                             & ~\cite{RodriguezICRC21} \\
&                          &                               & Andes$^a$                & Water Cherenkov               &                          & O(90\% fill-factor)           &                          & Underground                   & \\\midrule
SWGO    & Particle Sampler         & R\&D Phase                            & $>$ 4.4\,km a.s.l.       & Water Cherenkov                       & Inner $\approx$ 80,000\,m$^2$ & Inner $\sim$ 4\,m                & 8" PMT                   & WCD array$^d$                         & ~\cite{BarresICRC21} \\
&                          &                               & Andes$^a$                &                               & Outer $\approx$ 1\,km$^2$& Outer $\sim$ 16+\,m           &                          &                               & \\\midrule
& Air Cherenkov            &                               &                          & Timing array                  &                          & O(100\,m)                     & 8"/10" PMT               &                               & \\
TAIGA   & Air Cherenkov            & Pilot\,(1km$^2$)                      & Siberia$^a$              & Air-Imaging                           & up to 10\,km$^2$         & up to 600\,m                          & 3/4" PMT / SiPM          &                                       & ~\cite{Budnev2020} \\
& Particle sampler         &                               &                          & Scintillator SD/UD            &                          & $\approx$ 230\,m              & 1.2" PMT                 & Underground                   & \\ \midrule
\midrule
ASTRI   & Air Cherenkov            & Construction                          & 2.4\,km a.s.l.           & Imaging.                              & $\approx$ 40,000\,m$^2$  & 200\,m                                & 0.2$^\circ$ pixel-size SiPM  &                                   & ~\cite{AntonelliICRC21} \\
&                          & (est. 2024)                   & Tenerife                 & (dual-mirror)                 & 4\,m tel. $\diameter$      &                               & FoV $\sim$ 10$^\circ$    &                               & \\\midrule
CTA     & Air Cherenkov            & Construction                          & La Palma                 & Imaging                               & $\approx$ 1\,km$^2$, CTA-N & variable                            & 0.1$^\circ$ PMT (LST, MST) &                                     & ~\cite{ZaninICRC21} \\
&                          & (est. 2027)                   & Paranal                  & (multiple sizes)              & $\approx$ 10\,km$^2$, CTA-S &                            & 0.2$^\circ$ SiPM (SST)   &                               & \\\midrule
MACE    & Air Cherenkov            & Commissioning                         & 4.3\,km a.s.l.           & Imaging                               & Mono Telescope           &                                       & 0.12$^\circ$ pixel-size PMT  &                                   & ~\cite{YadavICRC21} \\ 
&                          &                               & Hanle                    &                               & 21\,m tel. $\diameter$     &                               & 1088 pixels              &                               & \\ \bottomrule
\end{tabular}
\begin{tablenotes}
\small
\item $^a$Evaluation for site of next stage currently ongoing.
\item $^b$Refers to average unit distances, which are arranged in regular clusters.
\item $^c$In a later configuration, WCDA-2 and WCDA-3 was equipped with 3"+20" PMTs, while WCD-1 remained with the orignal 1.5"+8" PMTs-configuration.
\item $^d$Options under investigation include double-layer WCD units~\cite{BiscontiICRC21} or multi-PMT WCD units~\cite{MercedesWCD}.
\end{tablenotes}
\label{tab:Summary}
\end{threeparttable}
\end{sidewaystable}

\vspace{0.5cm}
\noindent{\bf Particle Detector Arrays}

Particle sampling arrays measure the secondary air shower particles that reach the observation level. Many techniques can be applied to such purpose. The electrons and muons from an air shower are typically detected via their scintillation light produced by the ionization in scintillator detectors, or Cherenkov light produced, e.g. in water and ice. The produced light is detected using photomultiplier tubes (PMTs).
The light sensitive PMTs are connected to their respective light-producing medium in some light-tight container. From these basic working principles it results that the detector unit is insensitive to daylight and can be operated
continuously, with up to 100\,\% duty-cycle (as opposed to the $\sim$ 15\% of IACTs). 
A further advantage of particle arrays is a comparatively large field of view of typically 1\,sr (as compared to 0.024\,sr for a wide angle IACT with 10$^\circ$ field of view).
The large field of view and continuous operation duty cycle gives particle sampling arrays an advantage for extended sky surveys or  uninterrupted monitoring of sources over extended periods, without limitations due to daytime. 

Particle arrays use the arrival times and number of particles to reconstruct arrival direction, shower core impact position and energy. Limitations for the directional and energy reconstruction accuracies are the shower-to-shower fluctuations, and the width (arrival time width) of the particle shower front. This basic concept of sampling the shower front translates into two fundamental requirements, common to any technology applied: that of large active array areas (i.e., extended arrays with a good fraction of instrumented surface) and of high altitude installation sites, both necessary to achieve a satisfactory shower reconstruction and overall performance. Such detectors also have typically higher energy thresholds (in comparison to air-Cherenkov experiments), since only the most energetic showers penetrate deep enough in the atmosphere to produce measurable signals from charged particles or secondary high-energy photons at ground level.

The capability to discriminate between $\gamma$- and CR-induced air showers is another fundamental element of the technique, essential to achieve good sensitivity. Above several TeV, $\gamma$/hadron discrimination can be greatly improved by exploring the low muon content of $\gamma$-ray induced air-showers, using muon detection as a veto to supress the CR background. Placing particle detectors below ground or equipping them with shielding allows a measurement of the muon-component. This can also be used for a determination of the nature of the primary particle (cosmic ray composition).

At lower energies, cosmic-ray showers are muon-poor, and the muon cuts cease to be effective, so that $\gamma$/hadron discrimination must be based on the distribution of particles at observation level.
With a dense enough sampling of the footprint of the air shower (high array fill-factors), the comparatively irregular shape on the ground can then be used to separate hadrons from the more smoothly distributed particles in a $\gamma$-ray induce air shower.

For more on the particle detector arrays and water-Cherenkov technique see chapter "Particle Detector Arrays and Water Cherenkov Technique".

\vspace{0.5cm}
\noindent{\bf Air Cherenkov Technique}

Air Cherenkov detectors measure the Cherenkov photons emitted from the secondary EAS particles over the whole shower development, effectively using the Earth's atmosphere as a calorimeter, and achieve peak sensitivity between circa 100 GeV to few tens of TeV.
The advantage of the air-Cherenkov method is that the light can be detected over the full shower development, providing an effective calorimetric measurement of the energy deposited in the EAS. Due to the large number of photons emitted, of $O(10^5)$ for a 1 TeV $\gamma$-ray, energy resolutions of the order of 10\% are typically achieved. The air-Cherenkov pulses are short close to the shower core (order of 10\,ns), allowing to achieve a good angular resolution over a wide energy range. Furthermore, most of the emitted light in the optical range (mainly blue) reaches the ground with only little absorption, so that the energy threshold is lower as compared to the particle detection technique, where the air shower must have sufficient energy for the charged EAS particles to reach the observation level. Thanks to this low energy threshold, now approaching a few 10 GeV (in comparison to few 100 GeV for particle arrays), one advantage of air-Cherenkov instruments is to respond to transient alerts and follow-up variable sources. They nevertheless have a very limited duty cycle, operating only at dark time.

The air Cherenkov wave-front sampling technique was introduced by pioneering experiments such as (among others) THEMISTOCLE \cite{Baillon1993}, or AIROBICC \cite{Karle1995}. This technique provides an angle-integrating measurement of the light density on the ground, and the arrival times at individual detector stations, therefore being also referred to as a timing technique. It is the base for the next step by the TAIGA Collaboration, with the introduction of a hybrid reconstruction of air-shower data, using both IACTs and HiSCORE stations, in order to access the VHE and UHE $\gamma$-ray regime with a cost-effective instrumentation deployed over large detector areas.

In contrast to the angle integrating HiSCORE stations, IACTs provide an actual angular image of the air shower. At the core of the success of the technique is the efficacy of the imaging analysis in reconstructing the $\gamma$-ray shower and providing excellent hadron rejection~\cite{Holder21}. The Imaging Air Cherenkov technique was established with the first detection of the Crab Nebula by Whipple in 1989 \cite{Weekes1989}, and subsequent detections of several objects by, among others, Whipple, HEGRA and CAT. A further important innovation was the introduction of the stereoscopic observation technique by HEGRA, which is the widely used approach today. Here, multiple telescopes are arranged within an area about the size of the Cherenkov light pool in order to provide multiple simultaneous images of a same EAS event, from various viewing directions. Thanks to the use of stereoscopy, IACTs can achieve better shower core reconstruction, and as a consequence, better angular resolution ($\lesssim 0.1^{\circ}$) and hadron rejection than achievable in monoscopic observation mode using a single IACT. 

With the advent of the third generation of IACT experiments H.E.S.S., MAGIC and VERITAS, IACTs were established as the instrument of choice in the energy domain from 100\,GeV to several TeV. Thanks to their excellent sensitivity, which roughly scales with the number of telescopes in the array, IACT arrays are also good timing instruments. 

Among the key design characteristics of an IACT are its very large aperture, and mirror area, typically consisting of large (100+\,m$^2$) tesselated mirrors, which allow to collect as many Cherenkov photons as possible, and defining in turn the $\gamma$-ray energy threshold of the instrument. The telescope's large FoV, $\sim 3^{\circ}-10^{\circ}$, is also necessary to fully contain the Cherenkov images of the shower, a few degrees in angular extension, and usually offset by $\sim$ degree from the source position. This implies also the use of large, meter-sized cameras placed in the focal plane, and equipped with photomultiplier tubes (PMT). A good imaging of the air-shower requires a fine pixelation of the photosensitive area, meaning an array of hundreds or thousands of pixels with sizes of a fraction of a degree.This setup allows to image the fast air-Cherenkov pulses of EAS with great efficiency.

For details on the air Cherenkov technique, see chapter "The Air-Cherenkov Technique".


\section{TAIGA - Gamma-ray and Cosmic ray Astrophysics in Siberia}
\label{sec:taiga}
\subsection{The Tunka site}
%
The Tunka site is located in the Tunka valley in Siberia (51$^\circ$48'35'' N, 103$^\circ$04'02'' E) at an altitude of 675\,m a.s.l..
At these latitudes the temperatures during winter times can reach down to -50$^\circ$C. Astronomical observations are not possible during the short summer nights, mainly due to frequently arising thunderstorms, which put the instruments at risk, and only deployment operations are therefore concentrated on the summer months.

The Tunka site is hosting several experiments, some of which were initiated decades ago and continue to operate to this date, making up the valley's large complex.
Tunka-133 is the final stage of a cosmic ray experiment which has evolved from a smaller version in the early nineties, up to the current size of 175 optical stations distributed over an area of 3\,km$^2$ \citep{Berezhnev2011,Prosin2014}. 
The principle of operation of Tunka-133 is a measurement of the air Cherenkov light pulses from extended air showers (EAS) on the ground, using photomultiplier tubes (PMTs) pointing to the zenith. These measurements can only take place during dark clear nights.
Each Tunka-133 optical station consists of a hemispherical PMT (20\,cm cathode diameter, EMI 9359, Hamamatsu R1408) inside a metal cylinder. The dynamic range is increased to 3$\times$10$^4$, using one dynode and one anode readout channel.
The Tunka-133 array is organized in clusters of 7 hexagonally arranged detector stations, with a distance between stations of 85\,m. Tunka-133 consist of a core of 133 stations (19 clusters) covering an area of 1\,km$^2$, and 6 additional outer clusters, placed at a distance of 700-1000\,m from the center of the array, resulting in a total array area of 3\,km$^2$.

An array of scintillation detectors, Tunka-Grande, was installed in 2015, allowing to measure the muon-component of EASs \cite[e.g.,][]{Monkhoev2017}. Typically, hadronic air showers contain a factor 30 more muons than a gamma-ray air shower. Therefore, a measurement of the muon component can be very efficient for gamma-hadron separation. At the same time, this approach requires a large muon-active to total array area ratio, also referred to as filling factor. The Tunka-Grande detector component was the first step towards a particle detection array for TAIGA, as will be described below.

A radio extension, Tunka-Rex \cite{Bezyazeekov2015}, was part of the experiment until 2019, measuring the radio emission from EASs. Tunka-Rex consisted of 63 radio antennae, distributed over 3\,km$^2$ and was operated in coincidence with the Tunka-133 and Tunka-Grande arrays. As opposed to Cherenkov light measurements, the radio and particle detection techniques are not restricted to darktime. Finally, an optical telescope with 400\,mm diameter is operated on the Tunka site as part of the MASTER Global Network of Robot Telescopes \cite{Kornilov2012}.

The Tunka site offers valuable infrastructure and scientific environment for the development of the TAIGA experiment. The existing experience with different detection techniques and their operation under extreme conditions during the Siberian winter is of great benefit.

\subsection{Experimental Concept}
The Tunka Advanced Instrument for Cosmic ray and Gamma Ray Astronomy (TAIGA) is a hybrid detector concept for gamma-ray astronomy in the energy range from few TeV to several 100s of TeV, and for cosmic ray physics above 100\,TeV.

TAIGA emerged from a collaboration between University of Hamburg (UHH) and the Moscow state University (MSU), starting in 2009. At that time, the HiSCORE experiment only existed as a concept study under the name of SCORE \cite{Tluczykont2009,Hampf2009}. Due to the similar experimental approach of Tunka-133 and the excellent Tunka-site infrastructure, an agreement between MSU and UHH led to the first HiSCORE prototypes being deployed in the Tunka-valley in 2010. At the same time, first studies for a combination with IACTs were started. These activities eventually developed into the TAIGA collaboration, which today consists of 15 different institutions from Russia and Germany. TAIGA has deployed 120 HiSCORE stations, 2 imaging air Cherenkov telescopes (IACTs), and a first cluster of TAIGA-Muon detectors. 

All three components of the TAIGA experiment -- the air-Cherenkov timing array TAIGA-HiSCORE, the air-Cherenkov imaging telescopes TAIGA-IACT, and the particle detector array TAIGA-Muon -- measure the EAS on the ground, exploring both the secondary air shower particles, and the atmospheric Cherenkov photons to reconstruct the EAS properties.

The first component of the TAIGA detector complex was the wide-angle large-area wave-front sampling timing-array HiSCORE. The HiSCORE timing-array is based on the concept outlined in \citep{T2014}. This component started in 2010 as a stand alone concept, aiming at a cost-efficient coverage of very large detector areas in order to account for the steeply falling fluxes (power-law) with rising $\gamma$-ray energies. However, early on it became clear that while being a cost-efficient approach to cover large areas and achieve good core location reconstruction, and angular and energy resolutions, the HiSCORE concept alone suffers from poor $\gamma$/hadron separation power below 100\,TeV. Therefore, a combination with classical imaging air-Cherenkov Telescopes (IACTs), which provide good $\gamma$/hadron separation by using the EAS image shape, was envisaged. In order to cover large areas using IACTs in stereoscopic mode it is nevertheless necessary to position the telescopes at distances not more than 100-300\,m apart from each other. This is a significant limitation when aiming for very high energies, as it implies a large number of telescopes, and a large number of channels (PMTs \& electronics) per instrumented km$^2$.

Another option is to use the IACTs in monoscopic mode, placing them up to 600\,m apart from each other. While a standalone monoscopic IACT does not provide the same $\gamma$/hadron separation quality as stereoscopic systems, it was shown \citep{Kunnas2017} that a hybrid event reconstruction using the TAIGA-IACTs together with TAIGA-HiSCORE can achieve, at the same time, good $\gamma$/hadron separation and very large effective areas. Later on, an additional particle detector for the measurement of the muon component at higher energies was introduced, resulting in an improved hadron rejection at the high energy end.

\begin{figure}[t]
    \centering
    \includegraphics[scale=0.2]{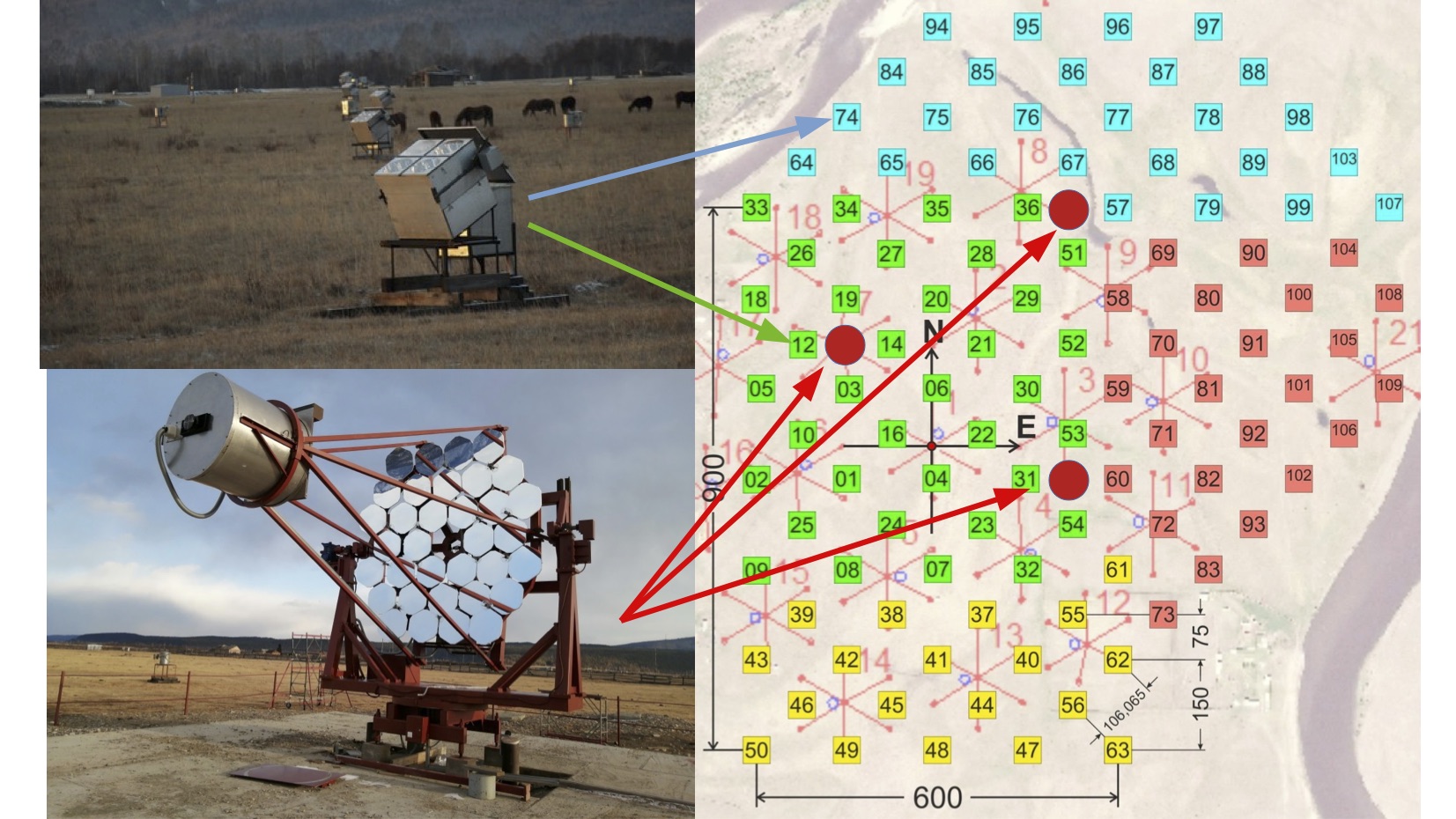}
    \caption{Layout of the TAIGA pilot array overlaid on a map of the Tunka observatory site. In total 120 HiSCORE stations (filled squares, 4 clusters in different colours) and 2 IACTs (red circles) are in operation. Clusters are defined as organisational units, the stations of one cluster all being connected to a central cluster controller.  By summer 2022, a third IACT will be installed.
    Adapted from \citep{Tluczykont2021}. Also shown are the detectors of the core of the Tunka-133 array (red dots, clusters indicated by lines).}
    \label{fig:TAIGA-layout}
\end{figure}

A general layout of the TAIGA-HiSCORE and TAIGA-IACT components on the Tunka site is shown in Figure~\ref{fig:TAIGA-layout}. In the following, the individual components of TAIGA are introduced, followed by a description of the hybrid reconstruction concept, and an overview of early results from the TAIGA pilot array.

\subsection{TAIGA-HiSCORE}

\vspace{0.3cm}
\indent\indent{\bf Station and array design}

The TAIGA-HiSCORE array currently consists of 120 angle-integrating air-Cherenkov timing detector stations, distributed over an area of approximately 1\,km$^2$.
The stations are arranged in an offset grid with distances between stations of 75\,m to 150\,m, as illustrated in Figure~\ref{fig:TAIGA-layout}.
The array is organized in 4 clusters comprising about 30 detector stations each, indicated by the differently coloured stations in the Figure. A cluster is an organisational unit with all stations of a cluster connected to a central cluster controller.
An individual station consists of 4 photomultiplier tubes (PMTs) equipped with a segmented Winston Cone. Both 8" and 10" PMTs from ElectronTubes and Hamamatsu are used. The Winston cones are built from light-weight reflective foil segments (Alanod 4300UP), and serve to reduce the background from stray light as well as to increase the light sensitive area to about 0.5\,m$^2$ per station.

At the altitude of the Tunka site, the resulting energy threshold is of 40\,TeV for gamma-rays, when using at least 3 stations for reconstruction.
The full opening angle of each cone is 60$^\circ$ wide, resulting in an effective field of view of 0.6\,sr (taking into account that the radial acceptance drops towards the edge of the FoV).
All stations are mounted on a steel construction which allows to tilt the optical axis along the north-south direction, therewith increasing the area of the sky that can be covered. Tilting all stations to the south increases the total field of view covered. Tilting all stations to the north concentrates the available observation time to a smaller total FoV, but with much deeper exposure \cite{Hampf2010}.
Currently, the HiSCORE stations are tilted by 25$^\circ$ to the south, in order to improve the total exposure on the Crab Nebula during the pilot phase of the TAIGA array.

\vspace{0.5cm}
{\bf Data acquisition and slow control electronics}

\begin{figure}[t]
    \centering
    \includegraphics[scale=0.4]{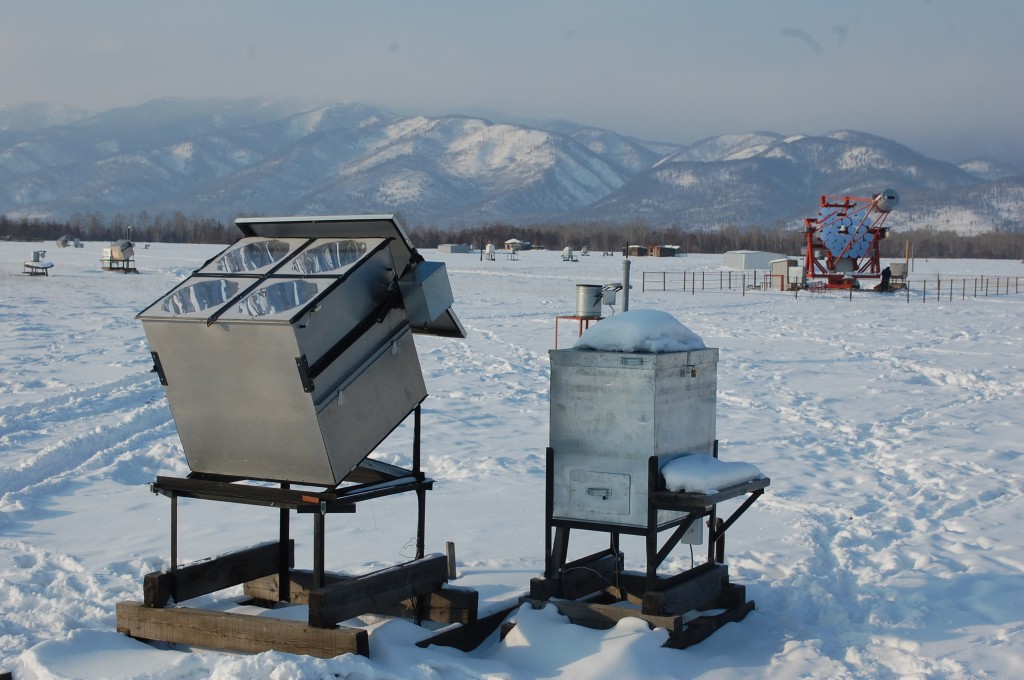}
    \caption{HiSCORE station (left) with electronic box (middle), and the first TAIGA-IACT (right, in the background). The HiSCORE stations are covered with a plexiglas window, equipped with a heating wire to avoid fogging. The electronics is placed in a separate, smaller box which is equipped with a stronger heating system, in order to heat up the components to at least 15\,deg Celsius before data acquisition is switched on. Due to the tilting of the stations, the entrance of the Winston cones can be seen. Figure: TAIGA Collaboration ({\it https://taiga-experiment.info/gallery}).}
    \label{fig:hiscore-station}
\end{figure}

One of the challenges of the TAIGA experiment is its location at high geographic latitude in Siberia, with harsh, cold winters.
The four cones of each station are insulated with foam and covered by a plexiglas window, which is equipped with a heating wire to prevent fogging.
The PMTs, placed at the base of the Winston cones, are equipped with divider bases, providing a nominal gain of 10$^4$ using 6 dynode stages.
In order to increase the dynamic range of the signal, the PMT anode and 5th dynode are read out.
The station metal box is equipped with a motorized lid, for protection against daylight and bad weather.

A separate, heated box, is installed next to each station and
contains the readout electronics and parts of the slow control and monitoring system.
The micro-controller based slow control system operates
the lid motors,
monitoring of lid motor and heating currents,
control of plexiglas window heating,
the high voltage (HV) control,
monitoring of anode current,
and implements safeguards in case of too high currents, or breaking daylight.

The PMT-station and electronics box are connected to the central DAQ via optical fibres for slow-control, readout, and distribution of a time synchronization signal. Additionally, a radio connection (XBeePro) is used for the heating controller of the electronics box. The DAQ system is not switched on before a temperature of 15$^\circ$C is reached inside the box.

\begin{figure}[t]
    \centering
    \includegraphics[scale=0.27]{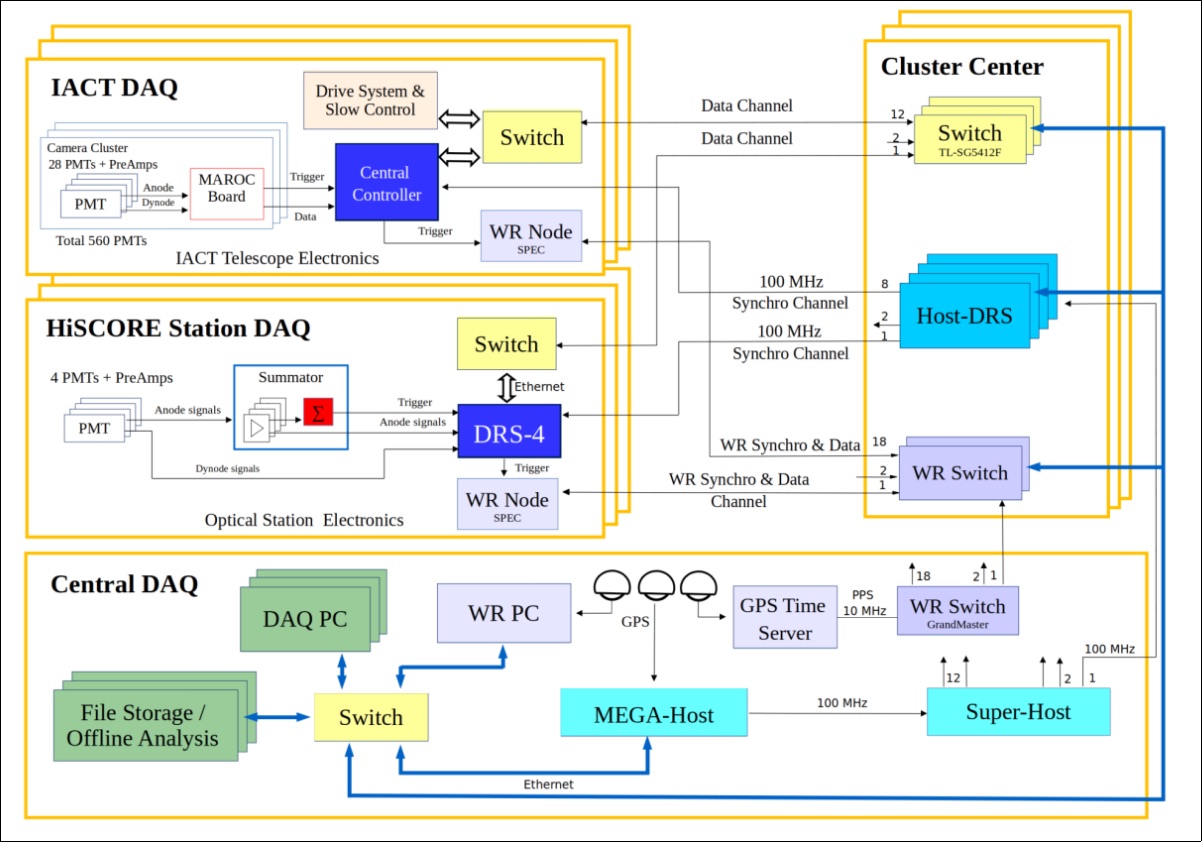}
    \caption{DAQ system of the TAIGA-HiSCORE array.
    The time-synchronization is performed using a distributed 100\,MHz clock signal. The WhiteRabbit system is also used in 20 stations for a cross check of the custom time synchronization. Block diagram from~\cite{Budnev2020}
    }
    \label{fig:hiscore_daq}
\end{figure}

The DAQ system of the HiSCORE array is illustrated in Figure~\ref{fig:hiscore_daq}.
The anode signals from the four PMTs are
connected to an analog summator and splitter board. The board output is an analog sum, used for triggering, and the four anode signals are then sent to the readout. The dynode signals are directly connected to the readout.
Triggering on the sum of four PMT anode signals reduces noise fluctuations, and therewith the threshold by a factor of 2.
If the analog sum exceeds a given threshold, all anode and dynode signals are read out using a DRS\,4 (Domino Ring Sampler \cite{drs4}) based readout board at a sampling frequency of 2\,GHz.
A 9th DRS\,4 channel is sampling a fiber-distributed 100\,MHz clock, allowing a relative time-synchronization between different HiSCORE stations. Of the 120 HiSCORE stations, 20 are additionally equipped with the WhiteRabbit ethernet-based time synchronization system which is used to cross check the DRS\,4-based time-synchronization \cite{Porelli2015}.
It was shown that a relative timing accuracy of 0.2\,ns over the full array is thus achieved \citep{Gress2017,Porelli2015}, which fulfills the requirement of a sub-ns time-resolution, necessary in order to reach a good angular resolution \cite{Hampf2013}.

A horizontal light source deflected into the stations is used for the measurement of the relative time-delays between stations. 
An estimation of the stability of the used time-synchronization systems resulted in an RMS value of less than 0.5\,ns for both systems \cite{Budnev2020}.
The detector stations are connected by optical fibre to their corresponding cluster center system.
The cluster centers are in turn connected to the central DAQ of the array.
Further details on the DAQ system and electronics components used can be found in \cite{Budnev2020}.

\vspace{0.5cm}
{\bf Data Reconstruction}

Each HiSCORE station triggers independently.
The event-building is done as a pre-processing step of the reconstruction at a later stage. The total amplitude, the time of the event (half-rise-time) and the signal width (full width at half maximum) are extracted for each detector. Based on these parameters, the air shower arrival direction, core impact position, and energy are reconstructed using methods developed ealier for Tunka-133 \citep{Berezhnev2011}, or HiSCORE itself \cite{Hampf2013}.

\begin{figure}[t]
    \centering
    \includegraphics[scale=0.4]{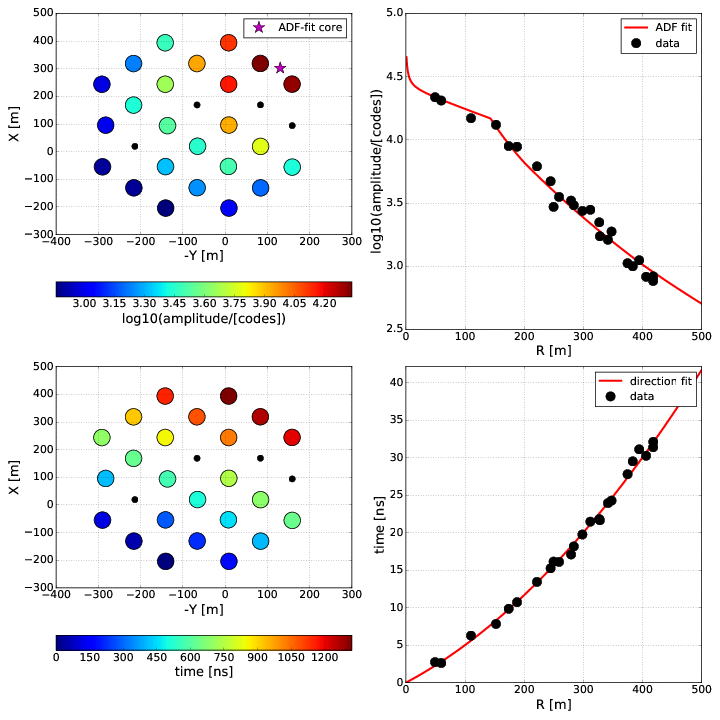}
    \caption{Real data air-shower event. Top left: distribution of amplitudes over the TAIGA-HiSCORE array. The position of the air-shower core impact position (star) is obtained from a fit of the amplitude density function (ADF) shown on the top right.
    The directional reconstruction is illustrated in the bottom: the distribution of arrival times on the ground (bottom left) is fit with an arrival time function (bottom right). From \cite{Porelli2020}.}
    \label{fig:event_display_hiscore}
\end{figure}

In a first reconstruction step, a 0th-order core position is estimated as the center-of-gravity of station amplitudes. Furthermore, a 0th-order angular direction is reconstructed using a fit of a plane to the arrival times. 
More precise core and angular reconstruction quality is achieved using models for the light distribution (light density function, LDF, and amplitude density function, ADF), and arrival time distribution on the ground. Figure~\ref{fig:event_display_hiscore} shows the distribution of amplitudes on the ground with the corresponding ADF fit as a function of core impact distance, and the distribution of arrival times at the stations with the corresponding arrival time function fit.The EAS energy is reconstructed using the value of the LDF at a fixed distance to the shower core (typically 200 m).

Using these methods, a reconstructed core impact resolution of 35\,m at threshold (only few stations used for fitting) and better than 10\,m at higher station multiplicities can be achieved.
As mentioned above, the angular resolution is better than 0.1$^\circ$. Finally, the energy resolution is of the order of 10\%, which is a typical value for air Cherenkov experiments.
The EAS maximum can be reconstructed using the slope of the LDF, the signal widths, signal rise times, or from an arrival time fit. However, the overall gamma-hadron separation of HiSCORE alone is poor at the threshold, and only reaches acceptable levels at several 100s of TeV (see, e.g. \cite{Hampf2013}).

During the commissioning phase of the HiSCORE engineering array (first 28 stations), a seredipitous discovery of a human light source was made \cite{Wischnewski2017}. Unexpected strong increases of the trigger rate during $\approx$\,1\,s intervals were found in the data set of the 2015-2016 season. This effect was found out to be due to a fast moving source close to zenith, which could be associated with the CATS-LIDAR (Cloud Aerosol Transport System) \cite{catslidar}.
The signal could be reconstructed using a plane wave fit and appeared 
as a point-like source for the HiSCORE array. The CATS-LIDAR was detected for 11 more passages in the subsequent observing season and provided a unique calibration tool for TAIGA-HiSCORE. During one of these passages, the robotic optical MASTER telescope was used to image the track of the source. These measurements were used to verify the absolute pointing of the TAIGA-HiSCORE array,
and to re-calibrate the relative time offsets between HiSCORE stations as well as their individual time jitter. It was also used
to estimate the angular resolution by comparison with the reconstruction of subsets of the array (chessboard method). This analysis resulted in a 0.1$^\circ$ angular resolution for plane-wave events.
Further details on this analysis can be found in \cite{Wischnewski2017}. 
The chessboard method had also been previously used in order to verify the TAIGA-HiSCORE performance in comparisons between data and MC simulations \cite{Epimakhov2015,Porelli2020}.

\vspace{0.5cm}
{\bf Monte Carlo Simulations and Array Performance}

Simulations of air showers were performed with the 
CORSIKA package \cite{Heck1998}. The detector simulation is done using an adaptation \cite{T2014,Hampf2013} of the sim\_telarray package \cite{Bernlohr2008} and a custom simulation chain \cite{Grinyuk2020,Postnikov2019}.
Both simulation chains implement the full detector response, including
(see, e.g. \cite{T2014})
Winston cone acceptance (based on ray-tracing simulations),
atmospheric photon scattering \cite{Kneizys1996},
wavelength-dependent PMT quantum efficiency,
photoelectron collection efficiency,
PMT signal pulse shape,
station trigger,
and PMT afterpulsing.

\begin{figure}[t]
    \centering
    \includegraphics[scale=0.12]{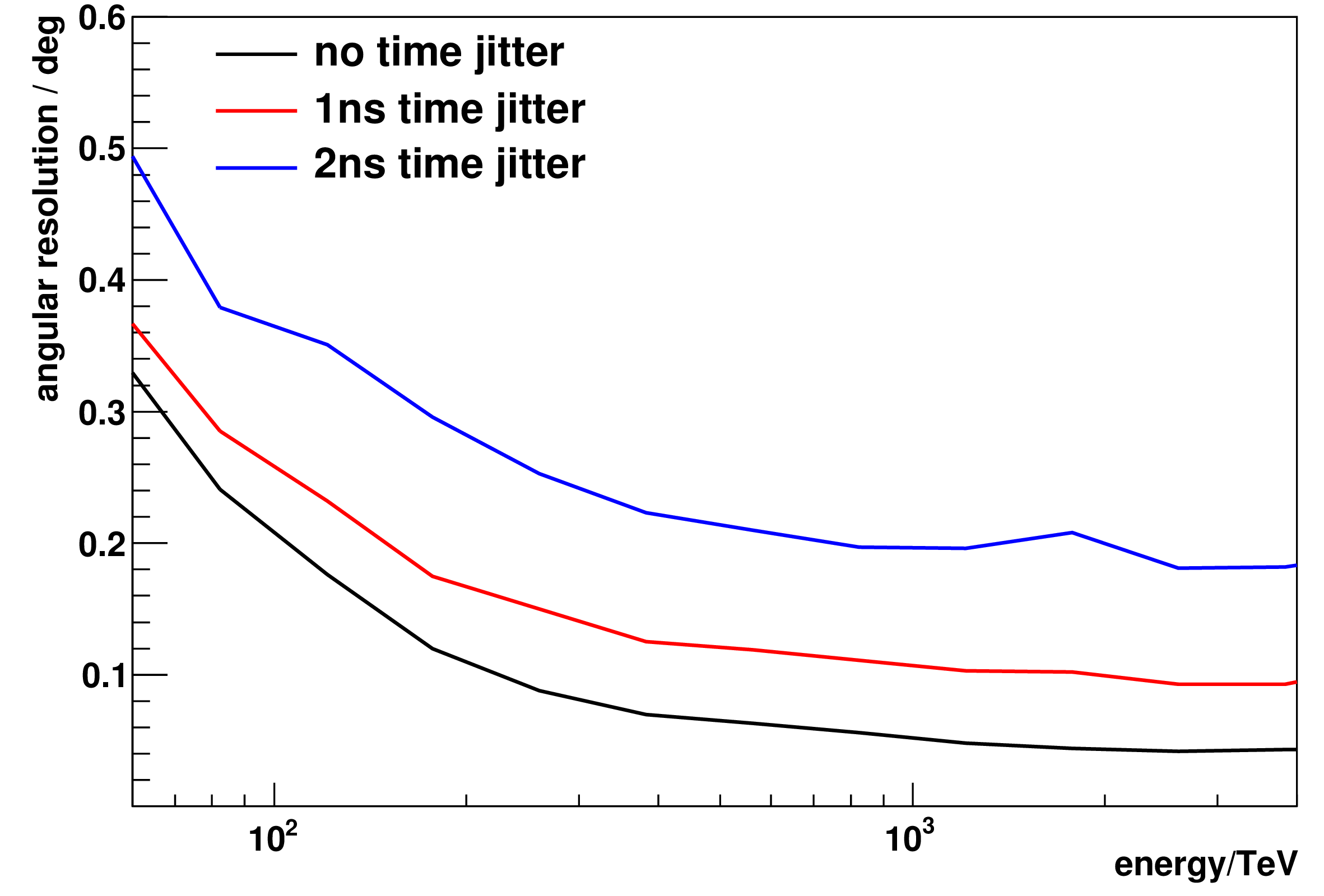}
    \caption{The simulated angular resolution of the HiSCORE timing array for different values of the relative time synchronization (here, labeled time jitter), adapted from \cite{Hampf2013}.}
    \label{fig:hiscore_angular_resolution}
\end{figure}

Monte Carlo simulations have shown that a sub-ns time-synchronization is required in order to reach the desired angular resolution of the order of 0.1$^\circ$. Figure~\ref{fig:hiscore_angular_resolution} shows the expected angular resolution using the reconstruction methods described above, assuming different relative time resolutions.
This predicted angular resolution could be verified using the above-mentioned analysis of the CATS-LIDAR, as well as through comparisons of real data to MC simulations using the chessboard method \cite{Epimakhov2015,Porelli2020}.

\begin{figure}
    \centering
    \includegraphics[scale=0.4]{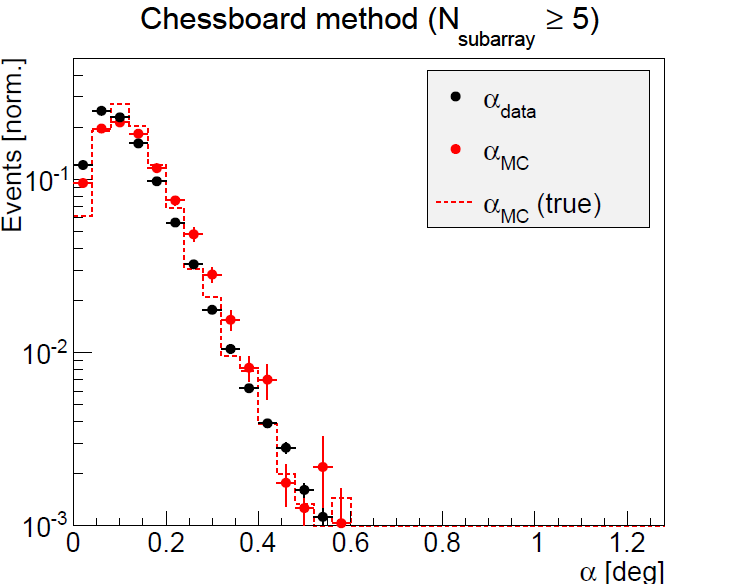}
    \caption{Using the chessboard method, a mismatch angle $\alpha$ between the reconstructed direction from two sub-arrays of the 28-station engineering array was calculated \cite{Porelli2020,Tluczykont2017}. The distribution shows that MC simulations reproduce the observed mismatch. This, however, does not show the angular resolution of the full array, because only few stations can be used in the subsets. MC simulations match the data well.}
    \label{fig:chessboard}
\end{figure}

The angular resolution is a function of the number of triggered stations.
When using small subsets of stations, the resulting mismatch, $\alpha$, between the reconstructed direction from both subsets is correlated to the angular resolution of the array. When using the same subsets in both data and MC simulations, the simulations can be verified, thus confirming the simulated angular resolution for the full array (shown in Figure~\ref{fig:hiscore_angular_resolution}), which goes down to 0.1$^\circ$ at energies above several 10s of TeV. The chessboard comparison is illustrated in Figure~\ref{fig:chessboard}, showing the mismatch angle $\alpha$ for a study done using the 28-station HiSCORE engineering array \cite{Tluczykont2017} (see also \cite{Porelli2020}).

The trigger rate of individual HiSCORE stations depends on the discriminator threshold used. The threshold was set to a value corresponding to 250 photoelectrons (p.e.), limiting the trigger rate to less than 20\,Hz. Simulations of this trigger setup and comparison to real data yield an energy threshold of 50\,TeV. This threshold might be further reduced in the future, provided a higher station trigger rate of the 120 HiSCORE stations can be handled by the DAQ system, possibly implementing online filtering methods. 

\subsection{TAIGA-IACT}

\vspace{0.3cm}
\indent\indent{\bf The IACT technique and TAIGA}


The driving idea of TAIGA is to access the energy regime up to several 100\,TeV. At these large energies, the amount of Cherenkov light produced in an air shower is very large, and small-size telescopes with diameters of 4\,m are sufficient. With a wide field of view of almost 10$^\circ$ diameter, air showers can be imaged up to core impact distances of around 500\,m from the telescope. When placing the TAIGA-IACTs 600\,m apart from each other, 4 IACTs are enough to cover an area of more than 1\,km$^2$. This is the opposite approach of classical stereoscopic systems, where the maximum distance between IACTs (100\,m to 300\,m, depending on the energy range) is dictated by the condition to cover each EAS with at least two telescopes.

\vspace{0.3cm}
{\bf TAIGA-IACT design}

\begin{figure}[t]
    \centering
    \includegraphics[scale=0.25]{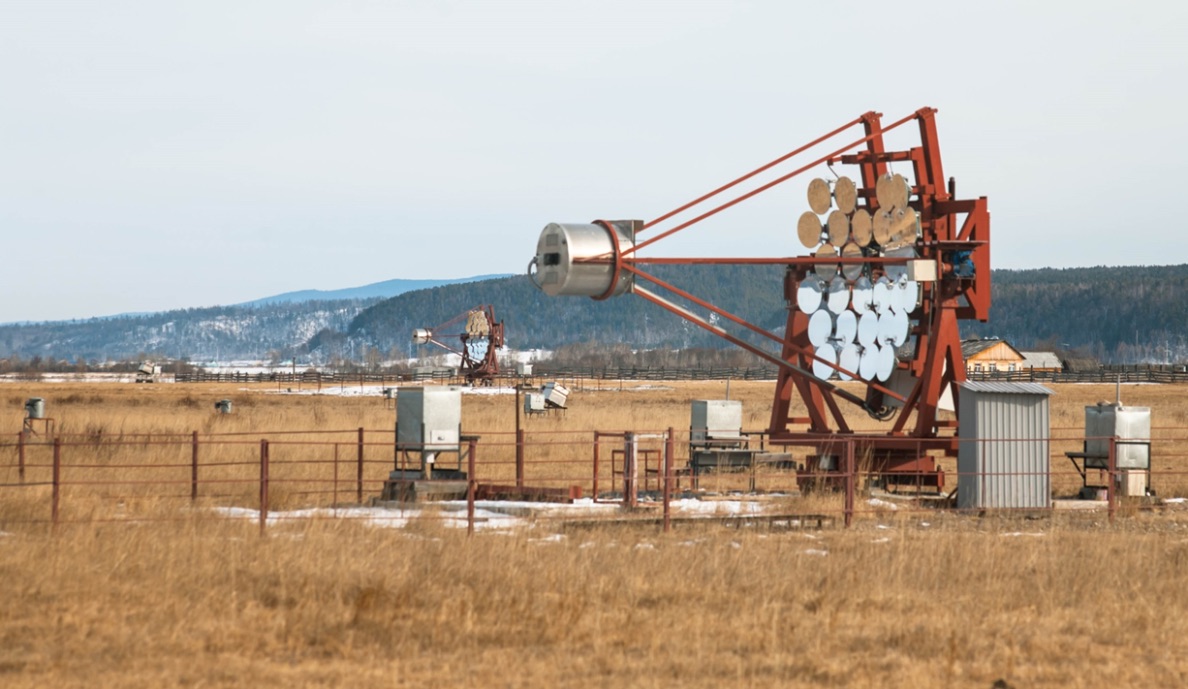}
    \caption{The first two TAIGA IACTs. At the time this picture was taken, the first IACT (foreground) was still equipped with round mirrors, later replaced by hexagonal mirrors. Picture: TAIGA-collaboration.}
    \label{fig:TAIGA-IACT}
\end{figure}

Today, three IACTs are in operation in TAIGA. 
Figure~\ref{fig:TAIGA-IACT} shows the first two TAIGA-IACTs.
The telescope dishes are built following the Davies-Cotton design, and are mounted on an elevation and azimuthal axis (alt-az). A mirror dish consists of 34 hexagonal mirrors with an effective diameter of slightly more than 60\,cm each, yielding a total effective light-collection area of more than 10\,m$^2$.
The focal length of the mirror dish is 4.75\,m, with an $f/d$ of 1.1.
The alt-az axes are driven by a Phytron hybrid stepper motor, equipped with 17-bit shaft encoders and stop switches. The telescope pointing is monitored and corrected using a sky-CCD camera system, imaging known bright stars.
Measurements of currents induced by stars drifting through the camera (tracking switched-off) were used to determine an absolute pointing accuracy of 0.02$^\circ$ \cite{Budnev2020} (also see \cite{Zhurov2019}).

A Cherenkov light camera is installed at the focus of the mirror dish.
The first IACT camera consists of 560 XP1911 PMTs, with a cathode diameter of 19\,mm. The second camera consists of 595 PMTs of the same type.
The PMTs are equipped with Winston cone light funnels, fabricated as a single plane. Each Winston cone covers the
full reflector diameter. With an angular diameter of 0.36$^\circ$ per camera pixel (PMT+cone), the full camera has a field of view of 9.6$^\circ$.
The camera body was especially designed with the harsh environmental conditions in mind, with insulated walls, a temperature control system, and a thick, 1.5 cm, plexiglas entrance window in front of the Winston cone plane. A camera lid protects the PMTs during daylight.

\begin{figure}[t]
    \centering
    \includegraphics[scale=0.3]{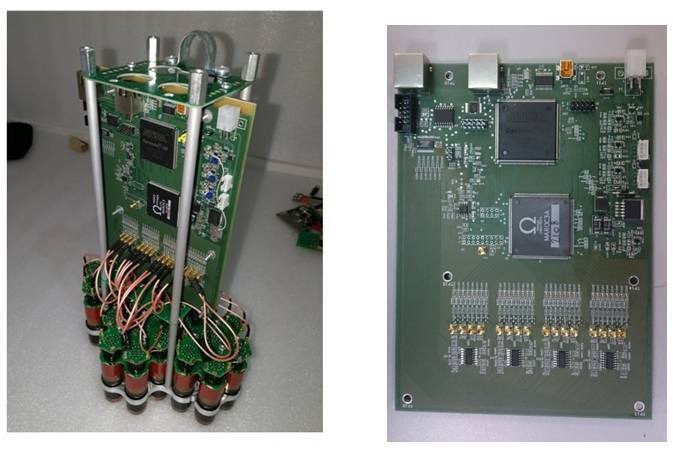}
    \caption{Left: Cluster of 28 PMTs. The PMTs of each cluster are read out by a MAROC-3 board (Right) \cite{Lubsandorzhiev2019,Budnev2020}.}
    \label{fig:TAIGA-IACT-sector}
\end{figure}

Each camera is organized in clusters of up to 28 PMTs (see Figure~\ref{fig:TAIGA-IACT-sector}).
The cameras have a field of view of about 9.6$^\circ$.
The high voltage is supplied in groups of 7 PMTs (up to 4 groups per cluster), selected with similar gains.
Each cluster includes HV control and current monitoring components.
Each cluster is read out with a board based on a 64-channel front-end ASIC, MAROC\,3 \citep{Blin2012}.
The trigger condition requires at least 2 pixels above a programmable threshold within one cluster.
A central camera controller, based on an FPGA Xilinx Spartan-6, is used to generate the global camera trigger and event time stamps, to manage the settings and readout of the MAROC\,3 boards, and for data transmission to the central DAQ.
The Central Controller includes a local clock, operated at a fiber-distributed 100\,MHz frequency from the DAQ center,
which is synchronized with all HiSCORE stations. 

For further details on the IACT design and electronics, see \cite{Yashin2016,Budnev2020}.

\vspace{0.5cm}
{\bf Event reconstruction}

When using a single IACT for gamma-ray observations, the gamma-hadron separation typically relies on a set of cuts, e.g. supercuts \cite{supercuts}, based on different image parameters, such as width or length, first introduced by \citep{Hillas1985}, and so-called Hillas parameters.
Mostly, two types of parameters can be used to cut on the direction of the events. The first parameter is the $\alpha$-angle, which is the angle between the major axis of the air shower image and the position of the observed source. The image major axis is pointing towards the direction of the EAS event. Therefore, in case of a $\gamma$-ray excess from the source position, an enhancement of events is expected at small values of the $\alpha$ angle. When using a single IACT, 
the classical approach is to use the angle $\alpha$ instead of a true directional information. However, it is also possible to directly reconstruct the actual direction $\theta$ using the $disp$-parameter.
This parameter is based on the ratio of image width, $w$, over image length, $l$. Using MC simulations, a relation between $w/l$ and the absolute distance from the center of gravity of the image to the position of the event direction along the image major axis inside the camera is generated.
MC simulations
for TAIGA show that this method can reach an angular resolution better than 0.2$^\circ$ in the energy range from few TeV to 10\,TeV.

\begin{figure}[t]
    \centering
    \includegraphics[scale=0.7]{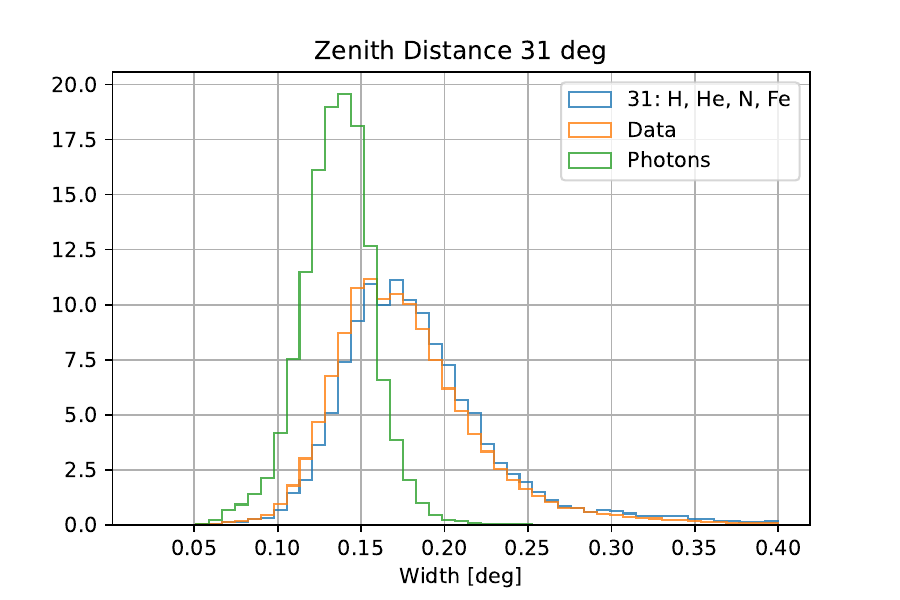}
    \caption{The width of air shower images obtained from MC simulations for gamma-rays (green) and hadrons (orange), compared to the observed image width of background data from the first TAIGA-IACT. From: \cite{Tluczykont2021}.}
    \label{fig:MC-comparison-width}
\end{figure}

While the goal of TAIGA is to implement a hybrid reconstruction using the IACTs together with the HiSCORE timing array, and the muon counters, an important step towards such a reconstruction is to verify the function of the IACTs using Monte Carlo (MC) simulations and observations of known sources.
Simulations of the TAIGA-IACTs are realized in two different MC chains. Both chains are using CORSIKA for air shower simulation \cite{Heck1998}. The detector simulation is done using an adaptation of the sim\_telarray package \cite{Bernlohr2008} (also see \cite{Kunnas2017,Blank2021}), using a custom simulation chain developed within TAIGA \cite{Postnikov2019,Grinyuk2020}.
First comparisons show a good agreement between data and MC. Figure~\ref{fig:MC-comparison-width}
shows the image's width parameter, obtained from the second moment of the image pixel distribution for background data of the first TAIGA-IACT, compared to the simulated width for hadrons and $\gamma$-rays.

Before implementing a full hybrid analysis 
the functionality of the TAIGA-IACTs must be demonstrated using real data.
The Crab Nebula is the standard candle of TeV $\gamma$-ray astronomy. Data on the Crab Nebula were taken during the commissioning phase of the first TAIGA-IACT and during subsequent observation seasons. 

\begin{figure}[t]
    \centering
    \includegraphics[scale=0.75]{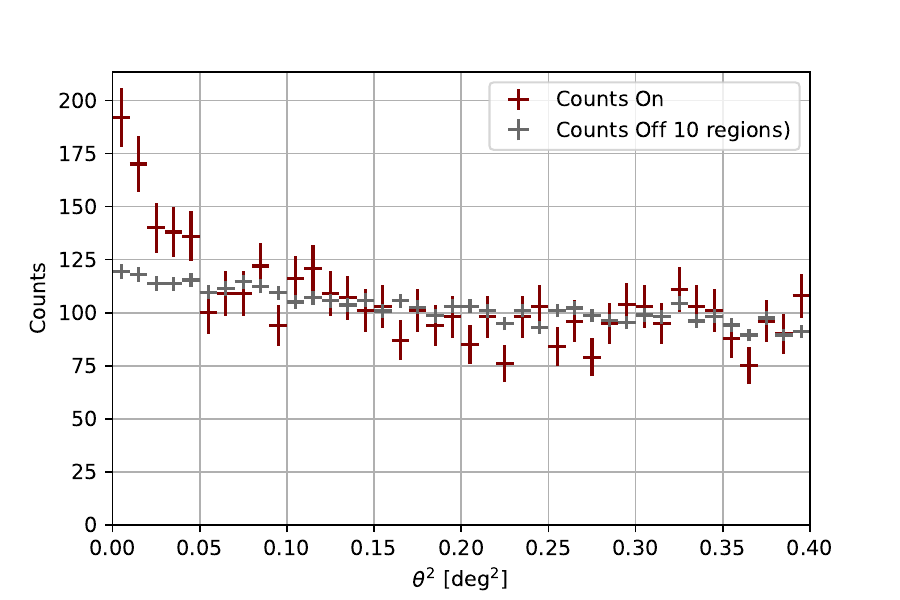}
    \caption{Distribution of the squared angular distance $\theta^2$ for Crab Nebula data of the first TAIGA IACT from the 2019/20 season. Courtesy \cite{Blank2023} (also see \cite{Blank2023MNRAS}).}
    \label{fig:theta_crab}
\end{figure}

A first analysis was based on 40\,h of good weather quality data under stable instrument conditions. Using a standard Hillas analysis with a cut on the $\alpha$ angle yielded 
an excess of 164 $\gamma$-ray events at a significance level of 6.3$\sigma$ \cite{Budnev2020}.
This result was confirmed by different groups using two independent reconstruction chains.
An analysis \cite{Blank2023,Blank2023MNRAS} using a MC simulation trained random forest algorithm based on a larger dataset of 80\,h, and using the disp parameter to reconstruct the direction of the primary gamma-rays, resulted in a larger significance (8\,$\sigma$ level) and a clear signal in the on-source (i.e. centered around the Crab Nebula) region at small values of the angular distance $\theta$, as shown in Figure~\ref{fig:theta_crab}. The average distribution of 10 off-source regions used for background estimation essentially remains flat towards low values of $\theta^2$.
The signal obtained from the Crab Nebula confirms the expected angular resolution at the current energy threshold of the first TAIGA-IACT
\citep{Blank2023,Astapov2022b}.

A drawback of the current TAIGA-IACTs is the large difference in field of view between the telescopes and TAIGA-HiSCORE. Only about 4\% of HiSCORE events fall into the FoV of the telescopes. Furthermore, the usage of Silicon-PMTs (SiPMs) has several advantages over the usage of standard PMTs, such as operation under full-moon conditions, no degradation due to high levels of background light, compact design, and low voltage and power consumption.
A small imaging telescope (SIT) prototype concept \cite{Chernov2020} addresses these points, based on a Schmidt optical system and a SiPM camera, providing a FoV of 20$^\circ$ in diameter. However, this study is in the prototyping stage, and a future use in the experiment is not decided.

\subsection{TAIGA-Muon}
An efficient $\gamma$/hadron separation is possible using the muon component of air showers at energies above 100\,TeV, when using an instrumented muon detector area of 0.2-0.3\% of the total TAIGA array (filling-factor). The goal is to build a TAIGA-muon array \cite{Budnev2020,Astapov2022a} of up to 3,000\,m$^2$.

In a first step, the Tunka-grande array consists of 19 scintillator stations. Each station has a surface and an underground component.
The surface component is built from 12, 80x80\,cm$^2$, scintillator tiles inside a protective steel-hut construction. The scintillator tiles were previously part of the EAS-TOP and KASCADE-Grande arrays. The underground component consists of 8 such detectors at a depth of 1.5\,m below ground. The stations are located at about 20\,m from individual Tunka-133 clusters, with distances between Tunka-Grande stations of 200\,m.

Based on the same concept, the TAIGA-Muon array \cite{Astapov2019,Astapov2022a} will consist of detector stations equipped with 16 counters, each consisting of 100x100\,cm$^2$ scintillator detector tiles \cite{Onuchin1992}. Four wavelength shifting light guides are used to guide the light of a counter to an EU-85-4 PMT. A TAIGA-Muon station will consist of 8 surface and 8 underground counters, yielding an area of 2$\times$7.52\,m$^2$.

\begin{table}[t]
    \centering
    \caption{Reconstruction parameters: classical stereo vs. hybrid.}
    \begin{tabular}{l|l|c|c}
                    & stereo  &  \multicolumn{2}{c}{hybrid} \\
                    &           & TAIGA-IACT      & TAIGA-HiSCORE \\\hline
         Direction  & axis intersection & $disp$-method & time-arrival fit \\
         Energy     & size      & size            & lateral density \\
         Core       & axis intersection & -- & lateral density \\\hline
    \end{tabular}
    \label{tab:hybridvsstereo}
\end{table}

\subsection{Hybrid imaging-timing concept}

For a single IACT, the $disp$-parameter yields an estimate of the position of the direction along the image major axis. However, this is only possible for a subset of events, because of an ambiguity between two possible positions at both sides of the elliptical shower image.
While higher camera resolutions can also help, stereoscopic IACT systems primarily resolve this ambiguity by using the intersection of major image axes of at least two IACTs. The same method is also used to reconstruct the EAS core impact position (with a projection into the observation plane) - see Table~\ref{tab:hybridvsstereo}. Stereoscopic IACT systems achieve angular and core position resolutions of typically 0.1$^\circ$, and better than 20\,m, respectively, on an event-by-event basis. The knowledge of the core position is key to an enhanced $\gamma$/hadron separation, based on the scaling of the image parameters (mostly width and length) with the values expected from simulations, and using the reconstructed core position, zenith angle and total image amplitude (size), measured in photo-electrons (p.e).

The caveat of stereoscopic observations is that at least two IACTs are needed to be located within the Cherenkov light-pool of the EAS event, thus small distances between IACTs are required (typically 100-300\,m, depending on the energy). Furthermore, stereoscopic events at far distances from the two IACTs are usually discarded due to very small angles between the intersecting image major axes (small stereo-angle). However, even though these requirements result in a smaller effective area than the sum of the effective areas of the individual IACTs - or expressed differently, in the required large number of channels per km$^2$ - in the past, the superior point-source sensitivity obtained by the very good angular resolution and $\gamma$-hadron sepration quality made stereoscopy the method of choice in air-Cherenkov astronomy.

\begin{figure}[t]
    \centering
    \parbox{\textwidth}{
    \parbox{0.65\textwidth}{
    \includegraphics[scale=0.13]{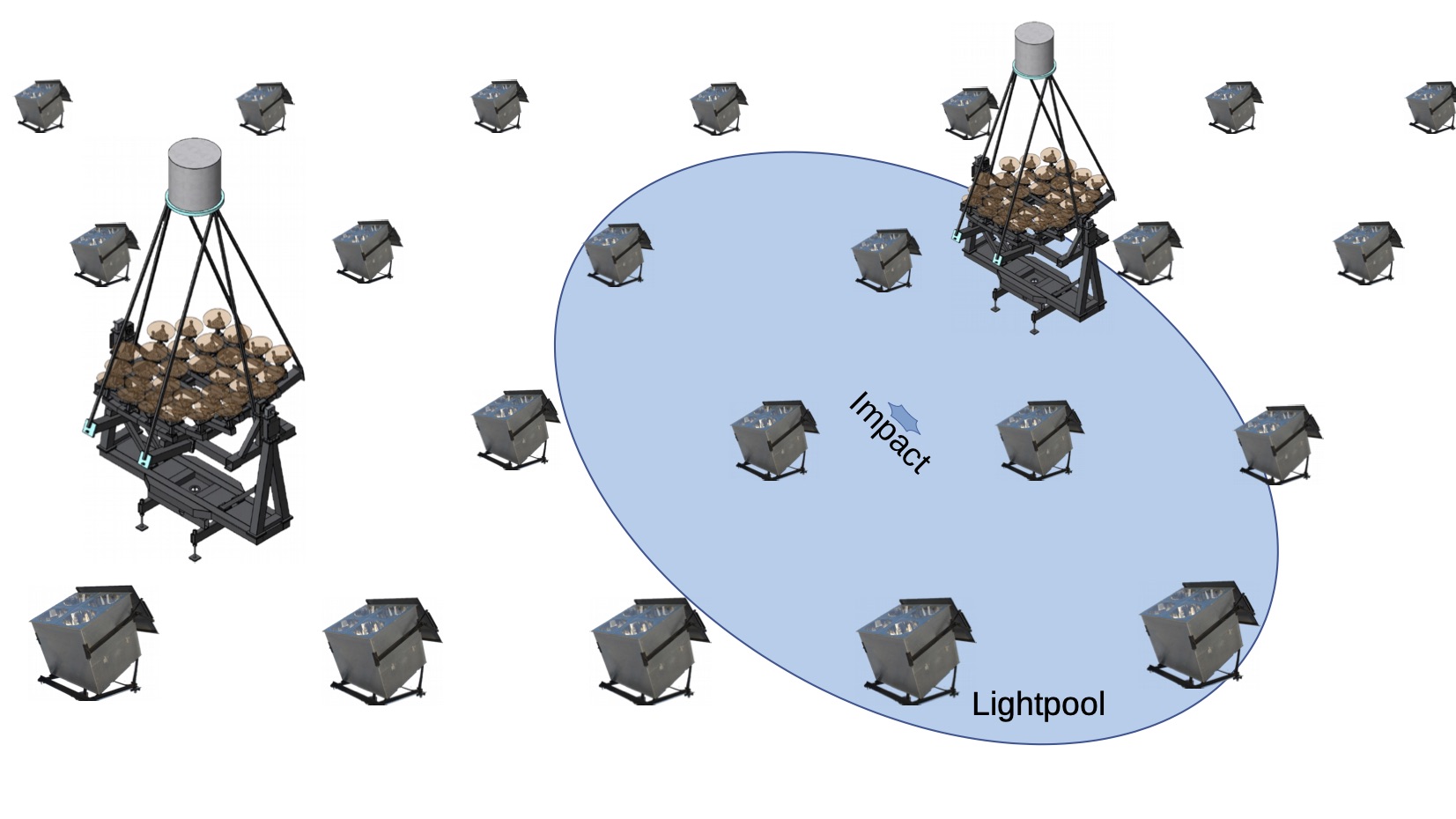}}
    \parbox[]{0.3\textwidth}{\includegraphics[scale=0.17]{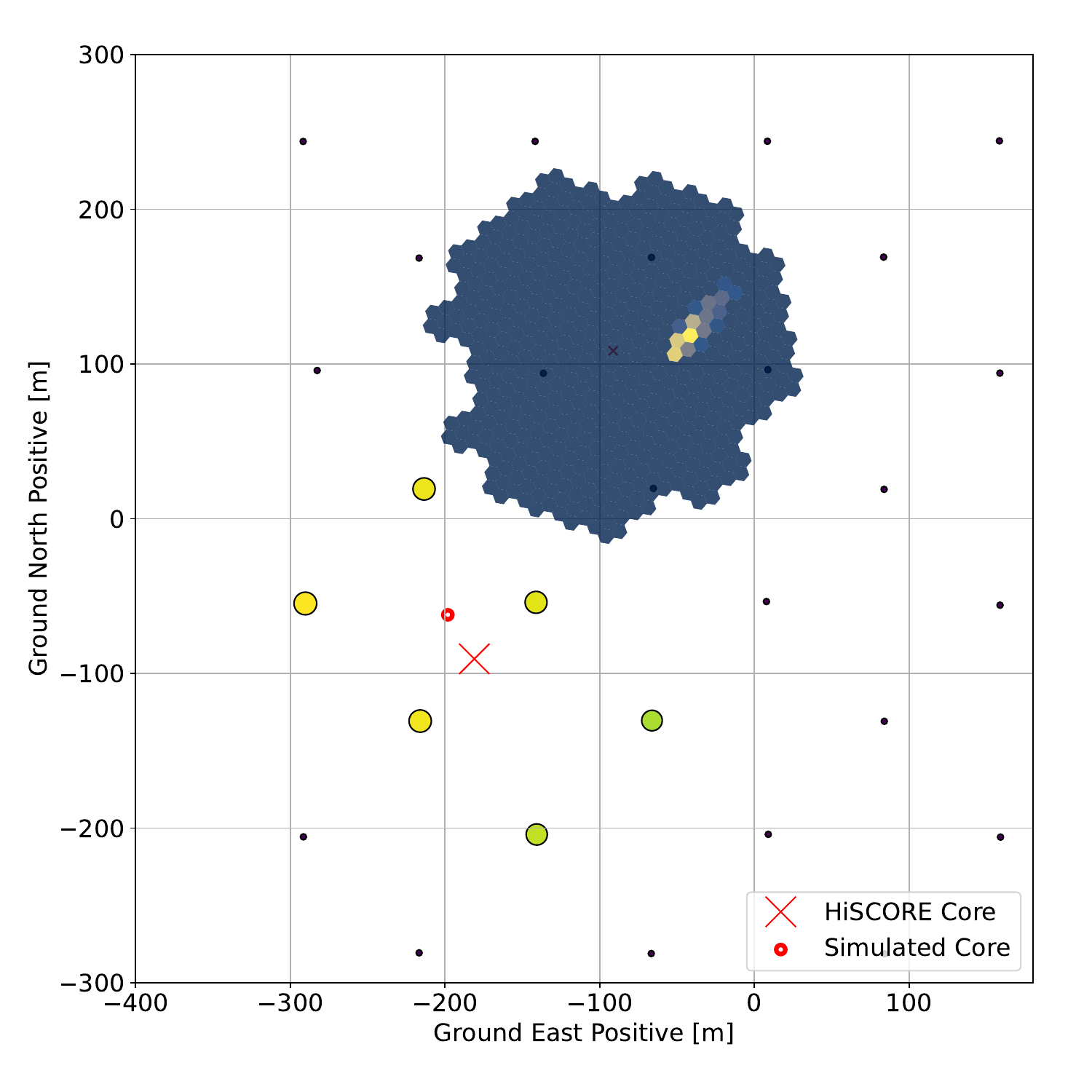}}
    }
    \caption{{\it Left}: schematics of the hybrid TAIGA principle. The TAIGA-IACTs will be operated at large distances from each other. The majority of events will be detected by a single IACT, and the surrounding HiSCORE stations, which provide the core impact position and direction of the air shower. {\it Right}:
    Amplitudes of HiSCORE stations and air shower image of the IACT corresponding to a simulated event illustrated on the left.
    Source: \cite{Blank2023}}
    \label{fig:hybrid_approach}
\end{figure}

The hybrid approach used within the TAIGA experiment is based on an idea to take advantage of the full available effective area of one IACT. This is achieved by placing the telescopes far apart from each other, so that most EASs will only trigger one single IACT, and several HiSCORE stations.
In order to compensate for the limitations of monoscopic IACT reconstruction, the telescopes are then combined with the HiSCORE timing array. While in the stereoscopic technique the EAS direction and core position are reconstructed using two IACTs, TAIGA uses the direction and core position as reconstructed by HiSCORE.
The principle is illustrated in Figure~\ref{fig:hybrid_approach}. The corresponding instrument response is also shown. It can be seen that the major image axis of the IACT projected to the observation level actually points in the direction of the core position reconstructed by HiSCORE. The latter is close to the simulated EAS core.

As illustrated in Figure~\ref{fig:hiscore_angular_resolution} for the angular resolution, the angular and core position resolutions of HiSCORE towards higher energies, as obtained from Monte Carlo simulations, are comparable to those of stereoscopic IACT systems \citep{Hampf2013}. The simulations were verified using data with a 9-station engineering array \cite{Epimakhov2015,Tluczykont2017,Porelli2020}.
Using the core position as reconstructed by HiSCORE, the zenith angle of the observation, as well as the image size measured by the TAIGA-IACT, the image width, $w$, can be scaled with its MC-expected value $w_{MC}$, thus obtaining the hybrid scaled width parameter ($hscw$).

\begin{equation}
    hscw = \frac{w}{w_{MC}(core,size,zenith)}
\end{equation}

\begin{figure}[t]
    \centering
    \includegraphics[scale=0.3]{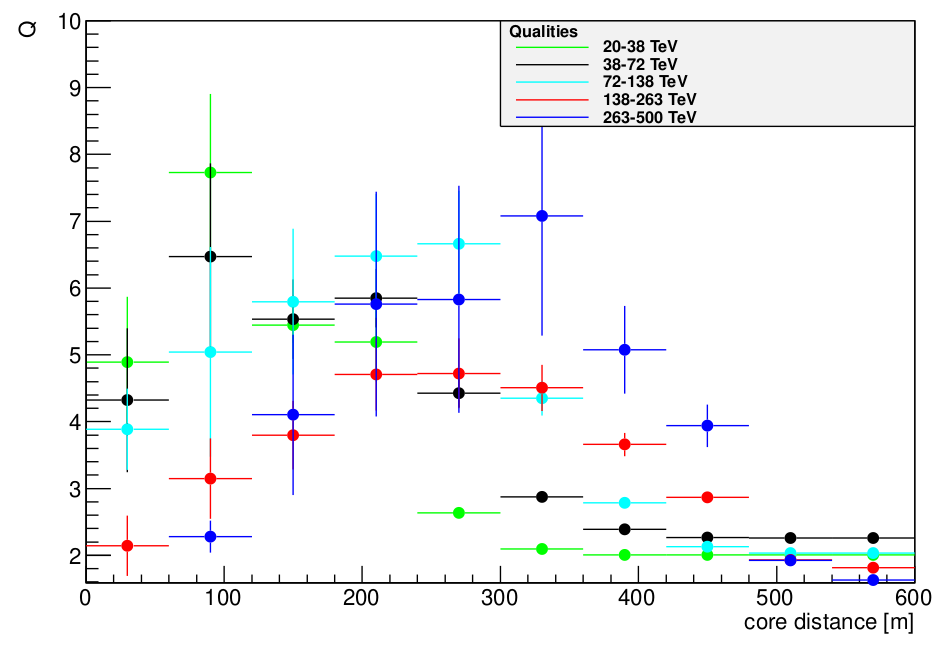}
    \caption{Quality factor Q of the $hscw$ parameter for the TAIGA hybrid reconstruction. Q is shown for different energy bins as a function of additional core distance cuts (rejecting events larger than the core distance on the x-axis).
    From \cite{Kunnas2017}.}
    \label{fig:kunnas-q-factor}
\end{figure}

Figure~\ref{fig:kunnas-q-factor} shows how the maximum quality factor of a cut on $hscw$, defined as the ratio of $\gamma$-ray efficiency to the square root of the hadron efficiency of the cut.

While for energies in the range from 20 to 38\,TeV the quality factor drops beyond core distances of 100\,m, the best quality factor for higher energies is reached at larger distances. A sweet spot, where the quality is optimized for all energy ranges considered, lies at core distances of about 250 -- 300\,m. Therefore, a distance between the position of two TAIGA-IACTs of 500 -- 600\,m is considered to be the best solution for the operation of the hybrid array. Integrating over all energies, and using a core distance cut of 250\,m (corresponding to a distance of 500\,m between IACTs), yields a very good overall quality factor of 4.6 for $\gamma$/hadron separation using only the $hscw$ parameter.

These results were obtained assuming a core impact resolution as obtained from simulations of the HiSCORE array \citep{Hampf2013}. It can be expected that a hybrid reconstruction, based on a combined fit of HiSCORE stations and IACTs will further improve the resolution, and thus the $\gamma$/hadron separation. Additional improvement of the $\gamma$/hadron separation can be obtained from other parameters such as the hybrid scaled-length, and the air-shower maximum reconstructed with HiSCORE,
or a fully combined hybrid fit.
Overall, an average energy-integrated quality factor better than 5 can be realistically expected.

So far, only events with a HiSCORE station multiplicity of at least 3 were considered for reconstruction. However, two-station events, or even single-station events, might in principle be reconstructed, provided an IACT also triggers the corresponding event. While the reconstruction quality will not reach the same level as achieved for higher energy events with higher station multiplicities, to recover such events will provide additional statistics at the threshold energy range of the HiSCORE array, around 10\,TeV. MC simulations yield a potential increase of more than 50\% statistics at these energies, when including these events.
While this is a special event class, so far not considered in the performance studies of TAIGA, it will be exploited in the future as part of the different types of event classes listed below.

\begin{enumerate}
    \item single IACT
    \item single IACT + 1-2 HiSCORE stations
    \item single IACT + 3 or more HiSCORE stations
    \item N IACTs + M HiSCORE stations
    \item N IACTs + M HiSCORE stations + K TAIGA-Muon stations at $E > 100$\,TeV
\end{enumerate}

The first event class will provide source monitoring in the range starting at the energy threshold of the IACTs, at few TeV. The second class can enhance the performance of the single IACT analysis in the energy range around 10\,TeV.
The third event class is aimed for the hybrid operation mode. Here, data from an IACT with core impact distances of up to 300\,m can be reconstructed, with an energy resolution of 10-20\,\%, and an angular resolution better than 0.2$^\circ$, over the full energy range from few TeV to few 100\,TeV; a $\gamma$/hadron separation with a quality factor of the order of 5 above 10\,TeV can also be achieved.
Event class number four will additinally provide stereoscopic events, which can be used to cross-check the hybrid reconstruction. In the current phase of the TAIGA experiment, the IACTs are placed closer together on purpose, in order to increase the fraction of stereo-events, for cross-check.
Finally, when adding the information of the Muon-content of the EAS as measured from TAIGA-Muon, the gamma-hadron separation at energies above 100\,TeV will be enhanced.

\vspace{0.3cm}
\indent{\bf TAIGA Sensitivity}

Sensitivities were estimated both for a pure HiSCORE setup \cite{T2014}, and for the hybrid Cherenkov technique \cite{Budnev2017}. Here, the difficulty lies in the fact that both detector components have a different field of view, different energy thresholds, and different operation modes. While the HiSCORE stations with their wide field of view of 60$^\circ$ in diameter are operated in scanning mode, oriented in a fixed direction\footnote{The optical axis of the stations are tilted to the south or to the north for longer periods of time (years), therewith accessing different parts of the sky}, the IACTs with a field of view of 10$^\circ$ in diameter are steered to point at any position accessible from the observation site.
Therefore, many sources will receive exposure time throughout the year with HiSCORE-only data, while some selected sources will be also observed with the TAIGA-IACTs.
Furthermore, the TAIGA-Muon detectors will be available at any time observations take place with HiSCORE or the IACTs.
Additionally, the TAIGA-Muon detectors will also operate alone during daytime.
With this in mind, it is clear that an estimation of the point-source sensitivity strongly depends on the part of the detector considered, and a straightforward comparison with other experiments is difficult.

\begin{figure}[t]
    \centering
    \includegraphics[scale=0.5]{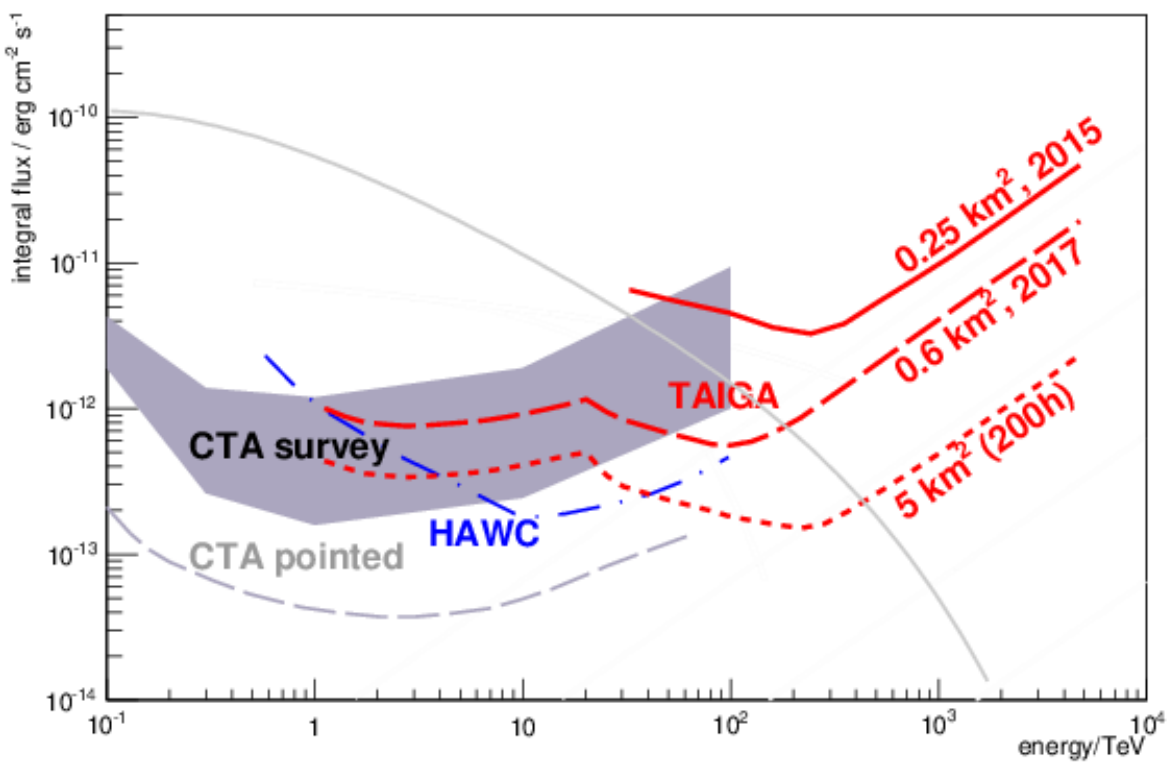} 
    \caption{Sensitivity of TAIGA using TAIGA-HiSCORE and TAIGA-IACT for different stages of the experiment (red curves). Currently, TAIGA has a size of 1\,km$^2$. The curves shown are based on a combination of simulations for the low energy part (IACTs) \cite{Kunnas2017} and a simulation of the HiSCORE component \cite{T2014}. Also shown are the point-source (50\,h) and survey sensitivities of CTA \cite{Dubus2013,Actis2011}, and the sensitivity of HAWC from \cite{Sinnis2005}. The grey solid line shows a model fit to Crab Nebula data \cite{Meyer2010}. Figure from: \cite{Tluczykont2017, Budnev2017}.}
    \label{fig:TAIGA-sensitivity}
\end{figure}

Figure~\ref{fig:TAIGA-sensitivity} shows the estimated sensitivity for the TAIGA air-Cherenkov detectors (TAIGA-HiSCORE and TAIGA-IACT).
At the low energy end, the sensitivity is dominated by the IACTs. No other detector component of TAIGA is sensitive at few TeV. Here, the first event class listed above will be collected.
In the energy range beyond 10\,TeV, TAIGA-HiSCORE starts to contribute to the reconstruction. Here, one- or two-station events in combination with an IACT can be used.
With rising energy, the number of stations will increase to more than 3 (event class 3). This is the main event class for TAIGA. In this range, the station multiplicity starts to be high enough to provide good angular and core position resolution, allowing a hybrid timing-imaging reconstruction.

Eventually the EAS will be large enough to also trigger a second IACT (event class 4). Such events can be used to cross-calibrate the hybrid event reconstruction, by using the classical stereoscopic technique. This is possible using the current setup, where the distances between the IACTs was chosen to be much smaller than 600\,m. However, in the future, the distances between two IACTs will be increased to about 600\,m, resulting in very few stereoscopic events at very high energies. 
Above 100\,TeV, the TAIGA-Muon component becomes relevant, since it will allow to measure the muon-content of EAS, providing additional hadron-tagging possibility. This improved hadron rejection was not taken into account in the sensitivity curve shown here. TAIGA-Muon is also operational in stand-alone mode during daytime, serving as cosmic ray detector.

TAIGA will also allow morphological studies with a good angular resolution ($\sim$ 0.1$^\circ$) and spectral energy reconstruction down to 10\% relative energy resolution. 

\subsection{Outlook}
After individual detector components have been verified, the next step is the implementation of a full hybrid reconstruction and a proof-of-principle of the hybrid concept within 2023.
The hybrid sensitivity of the current 1\,km$^2$ TAIGA array will reach down to few $\times$ 10$^{-13}$\,erg\,cm$^{-2}$\,s$^{-1}$.
In the future, a large 10\,km$^2$ array with order of 1,000 HiSCORE stations and 25 IACTs is envisaged.
Due to limitations of the Tunka site (total available area, weather quality), such a future site will be located elsewhere.
Such a future array will achieve a sensitivity of better than 10$^{-13}$\,erg\,cm$^{-2}$\,s$^{-1}$, at the same time allowing morphological studies and good spectral reconstruction. A modified design using either a combination of TAIGA-IACTs with the smaller sized SIT design described above might be envisaged.
Another possibility is to use variable HiSCORE station distances in a graded design, which could be shown to increase the effective area for a given number of stations while keeping a good angular reconstruction performance \cite{T2014}.

\section{Southern-Hemisphere EAS Array Proposals}

Until recently, the capabilities for effectively measuring $\gamma$-rays in the ultra-high-energy domain were relatively limited. Low $\gamma$-ray fluxes, coupled with high background levels from cosmic rays, make this a difficult regime for ground-based $\gamma$-ray astronomy, especially considering the technical challenges for achieving good $\gamma$/hadron separation. The perspective has nevertheless changed with the successful mapping of the Northern Hemisphere sky by HAWC above several TeVs~\cite{HAWC20} and the first firm PeVatron discoveries by LHAASO~\cite{ZhenCao21Nat}, which demonstrated that carefully designed particle array detectors, capable of circumventing the concealing backgrounds, can be effective instruments for UHE $\gamma$-ray astronomy. As all experiments in this energy domain have historically been located in the Northern Hemisphere, there is now considerable interest in developing facilities capable of measuring UHE $\gamma$-rays in the South, from where an extended view of the Milky Way, and access to the Galactic Center, store many anticipated discoveries. Additionally, in view of the new era of multi-messenger astronomy, and considering the potential of a wide-field array for monitoring the VHE transient sky, there are also strong reasons to try and reach better sensitivities at lower energies, below the current threshold of several hundred GeV. 

EAS detectors are now well-established, and have been proven ideal for spectral and morphological measurements of $\gamma$-ray signals above several tens of TeV. Future proposals are pushing the scientific frontiers of the technique, addressing a number of challenges driven by the need to achieve an improved sampling of the shower front as compared to previous experiments, and a wider dynamic range, at reasonable cost. When designing such facilities, elevations of circa 4\,km a.s.l. are sufficient for the highest energies, as UHE showers reach their maximum development at $\approx 4.3~\rm{km}\times log(log(E_0/1~\rm{PeV}))$. Higher sites could still represent an advantage if the goal is to lower the energy threshold, as the number of particles at ground level increases by several times from 4 to 5\,km a.s.l., for showers $\lesssim$ 1\,TeV~\cite{deNaurois-Mazin}. Higher altitudes can also contribute to improved energy and angular resolutions, as shower fluctuations are smaller at lower atmospheric depths, particularly near shower maximum. 

To fully exploit the benefits of high altitudes, and achieve the lowest possible energy detection thresholds, it is necessary that the spacing between detector units is reduced so to increase the array active area. To this purpose, a dense ($> 50\%$ fill factor) array core, large enough to at least contain the shower footprint ($\gtrsim 10^4~\rm{m}^2$), is to be applied. Such a large, dense array, installed at high altitude, can potentially achieve very good angular ($< 0.2^{\circ}$) and energy resolutions ($< 20 \%$) above few TeV. An extension of the dynamic range towards the PeV would depend on complementing the design with an outer array $O(10^6~\rm{m}^2)$. In the Southern Hemisphere, the Andes are the only suitable region for the installation of such arrays, given the availability of high-altitude plateaus.

 Looking into the different experimental options available, the water Cherenkov technique carries the advantage of being sensitive to the more numerous secondary $\gamma$-rays, which results in an improved sampling of the shower front, especially at larger distances from the shower core, where the ratio of secondary $\gamma$-rays to electrons further increases. This results in a potentially better angular resolution reconstruction of individual showers, by increasing the sampling rate of secondary particles and improving shape determination of the shower front\footnote{As will be shown, some experimental setups place a thin sheet of lead above conventional particle counters (e.g., scintillators) to yield additional signal from the conversion of secondary $\gamma$-rays into electrons that can be measured~\cite{Amenomori90}.}. Having a good angular resolution can also improve sensitivity, helping achieving a more favourable signal-to-noise ratio within the angular range of the point spread function. For that, in addition to a good sampling and shower core determination, accurate timing of the shower front at each detector unit is essential.
 
 In the following we will briefly present some of the current projects for Southern-Hemisphere ground-based gamma-ray detectors. The emphasis will be on the techniques proposed and how they differentiate with respect to the basic design considerations and the technology adopted. Although at very different stages of development and maturity, we find it useful to summarise and compare, whenever possible, the future performance expectations and potential of each proposal to advance the technology and measurement techniques.

\subsection{\textit{Southern Wide-Field Gamma-Ray Observatory, SWGO}}

The Southern Wide-Field Gamma-ray Observatory (SWGO) Collaboration was founded in July 2019 for the planning and design of a wide-field ground-based $\gamma$-ray observatory in South America. After the successful experience of the HAWC~\cite{HAWC17} water Cherenkov array, a few proposals emerged for the construction of the first Southern-Hemisphere installation of the kind, motivated by the vast scientific potential behind a continuous very-high energy survey of the southern sky~\cite{Mostafa17}. The initiatives were based upon the common concept of a dense, large-area, and high-altitude EAS array, which would significantly increase the VHE sensitivity over HAWC, especially towards the lower energies, below several hundred GeV. The SWGO Collaboration finally resulted from a joint effort between members of the SGSO Alliance\footnote{See the SGSO White-Paper here: \url{https://arxiv.org/pdf/1902.08429.pdf}.} and the LATTES Collaboration\footnote{Named after the Brazilian physicist Cesar Lattes, co-discoverer of the pion in 1947~\cite{Lattes}, and advanced mainly by groups in Brazil, Italy and Portugal, as well as the Czech Republic.}~\cite{LATTES18}. 

This initial proposal focused on a baseline design~\cite{Harm17} that would increase the observatory effective area over HAWC, and lower its detection energy threshold, by a combination of high array fill-factor (well-above 50\%, within a core area $\sim$ $10^{4-5}$\,m$^{2}$) and higher altitude installation site (close to 5\,km a.s.l.). Improved background rejection power by the efficient identification of muons at individual detector units is a central design goal. The concept also considers surrounding the central detector by a larger sparse array to provide higher energy sensitivity for the observation of PeVatrons.

\vspace{0.3cm}
{\bf The Observatory Concept}

Drawing from these ideas, the baseline concept for the SWGO observatory was defined as: 
\begin{itemize}
\item a ground-level particle detector array with close to 100\% duty cycle and order steradian field of view, to be installed in South America above 4.4\,km a.s.l, between latitudes -30$^{\circ}$ and -15$^{\circ}$.
\item to cover a wide energy range, from 100s\,GeV to 100s\,TeV, and possibly extending up to the PeV scale.
\item based primarily on water Cherenkov detector units, consisting on a high fill-factor core with area considerably larger than HAWC, and significantly better sensitivity, surrounded by a low-density outer array.
\end{itemize}

\begin{table}[t]
    \caption{SWGO baseline array configuration. Two WCD unit configuration options are considered.}
\begin{tabular}{p{2cm} p{2cm} p{7cm}} \toprule
    Component      & Parameter   & Reference design \\ \midrule
    Core Array     & Geometry    & 160 m radius circle = 80,400 m$^2$  \\
                   & Fill Factor & $\approx 80\%$, $\sim$ 5,700 units \\
    Outer Array    & Geometry    & at least 300 m outer radius = 202,200 m$^2$ \\
                   & Fill Factor & $\approx 5\%$, $\sim$ 880 units \\ 
    WCD units      & Double-layer & $\diameter$~3.8\,m; 0.5\,m (bottom layer) + 2.5\,m (top layer) height \\
                   & Multi-PMT  & $\diameter$~3.8\,m; 2.75\,m height \\
    Photodetectors & Option      & Large-area 8" PMT \\
                   & Geometry    & Central up/downward facing or 3-point star (\mbox{\large $\downY$}) \\
    Electronics    & Requirement & Nano-second inter-cell timing \\
    Reference Site & Altitude    & 4,700\,m a.sl. \\\bottomrule
\end{tabular}
\label{tab:SWGORC}
\end{table}

With respect to the development stage of the project, the SWGO Collaboration is in the R\&D Phase, which aims to deliver a detailed proposal that will guide the construction and operations of the future $\gamma$-ray observatory. The R\&D Phase is expected to be concluded in 2024, along with a final choice of the installation site, so that construction and operations could start as early as 2026. With a strong international collaboration of nearly 200 scientists from 14 member countries,\footnote{\url{www.swgo.org}} and a number of associated researchers from around the world, SWGO is the first global proposal for an EAS array in the Southern Hemisphere, complementing both LHAASO and CTA as the next-generation of ground-based gamma-ray observatories.

The SWGO R\&D programme is based on a Baseline Configuration which serves as reference for the array design and detector technology options to be investigated. Its main characteristics are described in Table~\ref{tab:SWGORC}. It consists of a core array of circa 5,700 water tanks, spaced in a grid with 4\,m gaps between units, and an outer array of at least 800 detectors, with an inter-unit spacing of 16\,m. 

As far as the WCD units are concerned, the same baseline design is considered for the core and the outer arrays, and two major design options are under study. The first consists of a double-layered (2.5\,m height top; 0.5\,m height bottom) cylindrical tank, with a diameter of 3.8\,m~\cite{Bisconti2022, Kunwar22}. Calorimetry of the shower electromagnetic component is done in the upper WCD layer, whereas the bottom layer is primarily for muon tagging. Each layer is equipped with a single, large-area photo-multiplier tube (PMT) placed at the center. Deployment in a lake for improved shielding from lateral penetrating particles is being investigated. The second option is a multi-PMT shallow-WCD tank with a diameter of 3.8\,m and 1.75\,m height, which aims to explore the asymmetric illumination of three upward-facing PMTs to identify individual muons~\cite{MercedesWCD}.

The key science topics which the SWGO Collaboration aims to address, some of which are unique to southern-hemisphere installations, include:
\begin{itemize}
\item At the lowest energies, $< 1$\,TeV, focus is on transient sources, exploring the wide-FoV and near-continuous duty cycle of the observatory to work as a monitoring and trigger instrument complementary to CTA. The principal science targets are Active Galactic Nuclei (AGN) and Gamma-ray Bursts (GRB), both of which are candidate multi-messenger counterparts of VHE neutrinos and gravitational waves, respectively. 
\item At the other extreme of the energy range, $> 100$\,TeV, science questions are dominated by the search for PeVatrons, putative sources responsible for the acceleration of knee cosmic-ray particles.
\item Access to the Galactic Center and Halo offer the possibility of in-depth searches for Dark Matter signals up to $\sim 100$ TeV, constraining therefore the entire energy range of WIMP models.
\item The study of Galactic diffuse emission, and extended sources, such as PWNe and TeV Halos, are an important target which will benefit from improved angular resolution at energies above several 10s\,TeV.
\item Finally, efficient single-muon detection capability will allow precise measurement of the muon content in hadronic showers, opening the way for mass-resolved cosmic-ray studies from 10s\,TeV to the PeV scale.
\end{itemize}

In the following we will detail some of the array configuration and experimental detector work underway for the observatory design.

\vspace{0.3cm}
{\bf The Array Configuration Evaluation}

\begin{figure}[t]
    \centering
    \includegraphics[width=\textwidth]{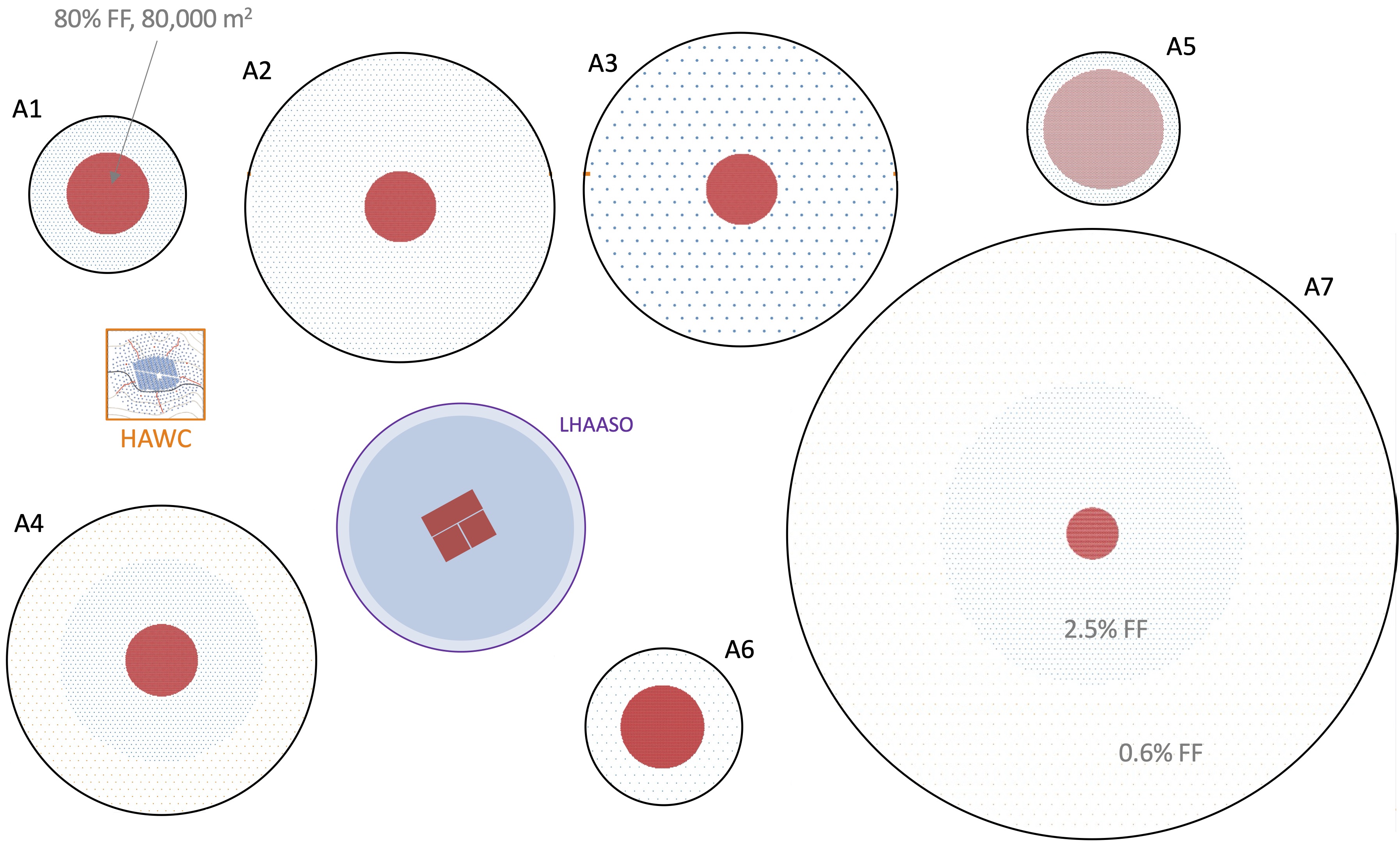}
    \caption{Illustration of the array configuration options currently under investigation by the SWGO Collaboration. The baseline array configuration described in Table~\ref{tab:SWGORC} refers to option A1. Image Credits: The SWGO Collaboration.}
    \label{fig:SWGOArrayConfigs}
\end{figure}

The Observatory layout will consist of a compact core surrounded by a sparse array. The SWGO Collaboration is currently investigating the array configuration options (see Figure~\ref{fig:SWGOArrayConfigs}) that impact performance on the basis of a predefined set of quantitative science benchmarks~\cite{BarresICRC21}. The basic layout parameters to be investigated are the total array area and fill-factor, as well as the site elevation, which will primarily impact the energy detection threshold. In assessing the array sensitivity, the main performance figures will be the effective area\footnote{Defined as the geometrical array area convolved with the detection efficiency.}, the $\gamma$/hadron discrimination efficiency, and the angular resolution, all considered over a target energy range. Ultimately, these parameters will reflect the fraction of the shower energy registered by the array, and in consequence the amount of information available for shower reconstruction (see e.g.~\cite{Hofmann21}). 

The main array configuration trade-offs (at a fixed cost) will likely play out between the low energies ($<$ 1 TeV), mainly driven by the site elevation, fill factor, and detector unit threshold, and the high energies ($> 100$ TeV), driven by the overall array area and background rejection efficiency. In addition to the array layout, the design of the individual WCD stations (including geometry, size and choice of electronics and photosensors) will determine the energy threshold to secondary particles, dynamic range and timing accuracy of the detector units, and will be discussed further ahead. The ability to discriminate between $\gamma$- and cosmic ray-initiated air showers is another fundamental aspect of the design. 

For the low energies, a dense, and sufficiently large core array is fundamental, in order to achieve good sampling of the entire shower front, of $O(10^4~\rm{m}^2)$. Here, background discrimination is mostly based on the different ground patterns of active stations produced by hadronic and $\gamma$-induced showers, exploring the differences between the lateral distribution of particles with respect to the shower core position. Critical design factors for the low-energies are the amount of signal collected, and the massively increasing background shower rates, 
which require a suitable trigger strategy to guarantee that an effective reduction of the energy detection threshold is achievable. Large fluctuations in the development and ground signal from low energy showers also play a critical role, usually destroying energy resolution, and altitude is a critical parameter. 

The angular resolution of an EAS array depends mainly on the array size and the temporal resolution of single stations, which is typically of $\sim \rm{few}$ ns for the WCD stations considered for SWGO. A high fill-factor core will also impact the precision that can be achieved in the shower core reconstruction. In the low energy regime, the sensitivity of the WCD stations to all shower particles, including secondary photons, which are $ \sim$ 10$\times$ more numerous than the charged e$^-$e$^+$ pairs, represents an important design advantage, which contributes to improving angular resolution. The energy resolution, obtained from fitting the lateral signal distributions at ground as a function of shower core distance, also benefits from all-particle sensitivity. Nevertheless, because of greater shower fluctuations, it is inevitable that the angular and energy resolutions will degrade significantly at low energies. A worse angular resolution will have, in turn, a negative effect in the signal-to-background ratio $N_{\gamma}/\sqrt{N_{\rm{CR}}}$, and consequently sensitivity\footnote{The number of background events is $N_{\rm{CR}}(E) = \Delta\theta^{2}\epsilon_{\rm{rej}}N_{\rm{CR}(E)}$, where $\Delta\theta^{2}$ is the angular resolution and $\epsilon_{\rm{rej}}$ the CR rejection efficiency, both of which severely degrade below $\sim$ 1 TeV.}. 

For the highest energy $\gamma$-rays, a sparse outer array is a cost-effective way of improving effective area, as the particle density in energetic showers allows for a good reconstruction with fewer stations spread over a larger area (fill-factors of few \%). The outer array also plays an important role in increasing the muon sensitive area, as these particles tend to have higher transverse momentum and spread to larger ($> 150$ m) distances from the shower core, over an area $\sim 10^5$ m$^2$. The critical array design factor is therefore to define the minimal density of stations needed for a good sampling of the shower front and guarantee the desired $\gamma$-hadron discrimination, shower core localisation, and energy resolution.  

The fact that cosmic-ray showers are about 4 orders of magnitude more abundant than $\gamma$-rays at these high energies imply that hadron rejection levels of 10$^{-4}$ or higher are required. Fortunately, the number of active stations hit by muons will be high enough so that excellent $\gamma$-hadron separation can be achieved provided that high muon detection efficiency, as well as the required array areas, are available. In principle, the large amount of electromagnetic energy deposited in the stations (mostly from secondary high-energy photons from $\pi^0$ decays in proton events) can also be explored for hadronic rejection. The requirements for $\gamma$-hadron discrimination, observatory sensitivity and energy/angular resolution, will ultimately drive the trade-off (at fixed cost) between array density and total area covered for the outer array. 

\begin{figure}[t]
    \centering
    \includegraphics[width=\textwidth]{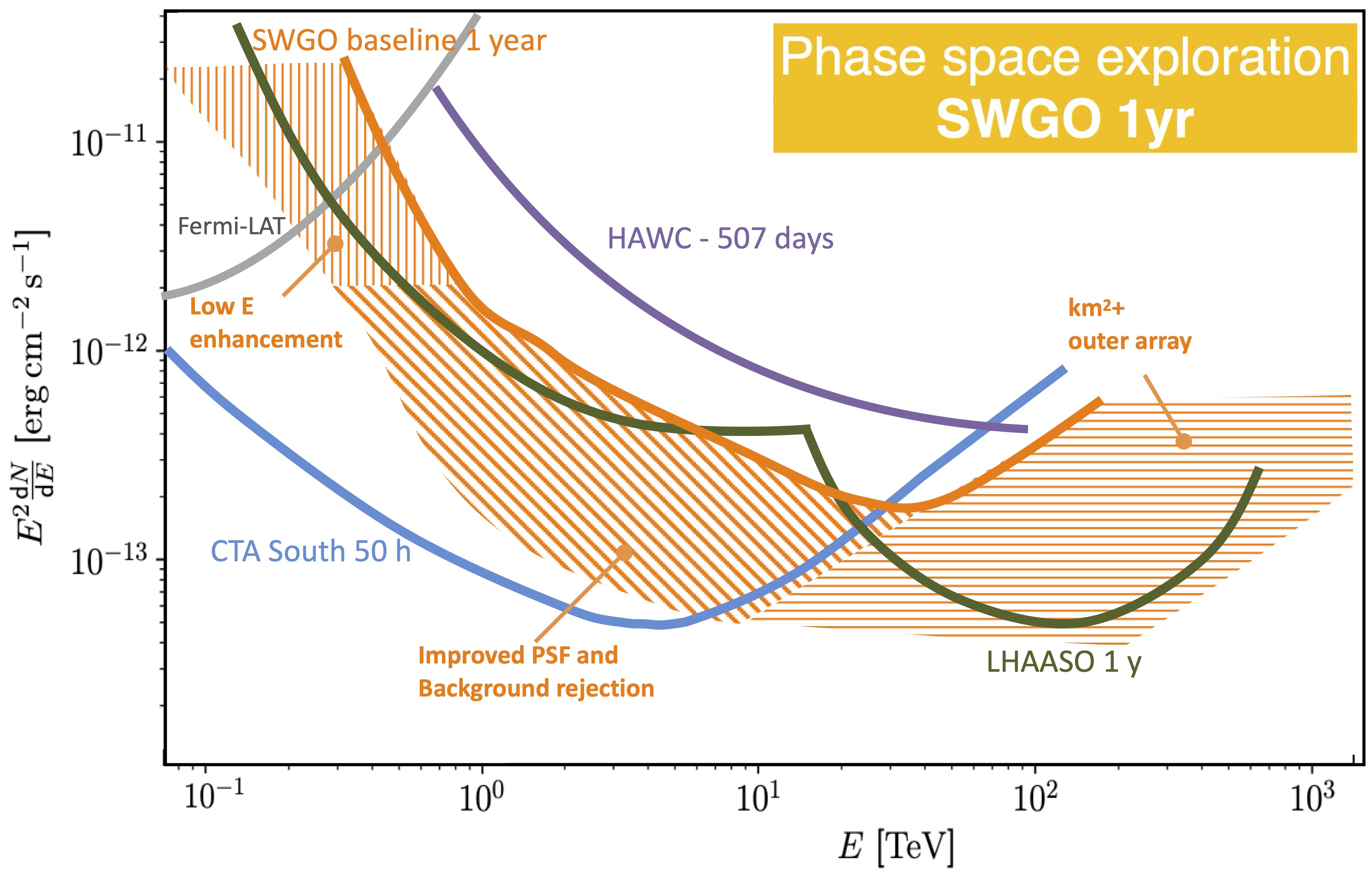}
    \caption{Phase-space explored in the design studies for SWGO. The orange bracketed phase-space under consideration is compared to the differential point-source sensitivity of various experiments~\cite{HarmICRC21}. The \emph{"baseline"} curve refers to the configuration sensitivity presented in~\cite{Harm17}, equivalent to array A1. The lower limit of the yellow band corresponds to a factor of 30\% improvement in the PSF and a factor of 10 improvement in background rejection efficiency. The size of the outer array is the main parameter driving the high-energy enhancement.}
    \label{fig:SWGOsensitivity}
\end{figure}

Current simulation work is being conducted for evaluating the performance of various array configurations and detector choices~\cite{HarmICRC21}. Figure~\ref{fig:SWGOsensitivity} shows the phase-space under exploration in the R\&D Phase, bracketed by the array options and detector unit designs under consideration. The performance baseline is set by the minimum configuration described in Table~\ref{tab:SWGORC} -- array configuration A1 in Figure~\ref{fig:SWGOArrayConfigs}. In general terms, a lower $\gamma$-ray energy detection threshold can be achieved by reducing the individual unit threshold and deployment at higher elevation sites. Improvements in angular resolution and background rejection will result in overall sensitivity gains. The higher energy enhancement indicated above 100\,TeV will be driven by the size of the outer array and background rejection efficiency, which shall scale with the total muon detection area available. 

The optimisation work is carried out for a same observatory location and magnetic field, at predefined altitudes between 4.1-5.2 km, as well as for a fixed estimated total cost.

\vspace{0.3cm}
{\bf The Detector Design Options}

The individual detector units define the accuracy of local measurements of the arrival time and energy density of the shower particles, as well as the capability for single-muon identification, and will directly impact the overall array performance. The large scale of the observatory, and the altitude and remoteness of the installation site, imply the need for little or no maintenance as a major design goal. Water scarcity and environmental concerns also present important constraints.

\begin{figure}[t]
    \centering
    \includegraphics[width=\textwidth]{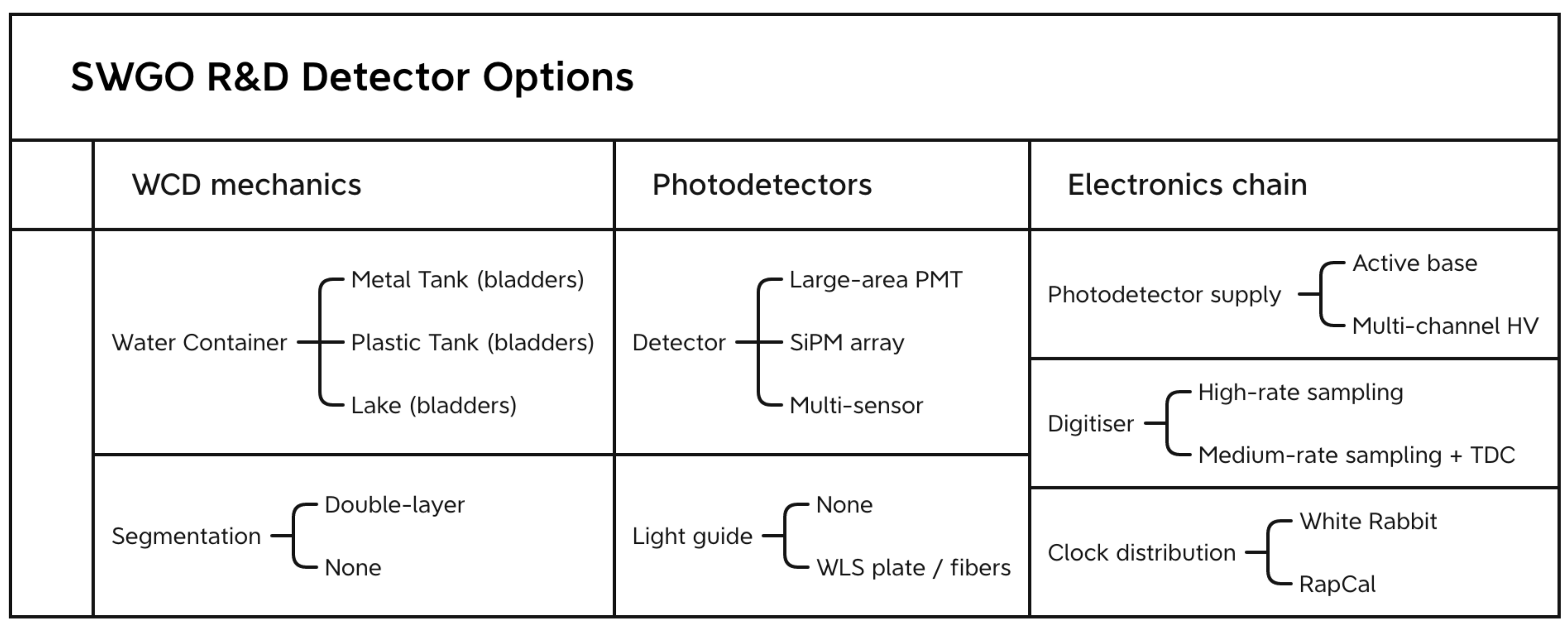}
    \caption{Decomposition of SWGO detector components into main options under study~\cite{WernerICRC21}.}
    \label{fig:SWGOoptionstable}
\end{figure}

A number of technological options are under investigation for the individual WCD units, as presented in Figure~\ref{fig:SWGOoptionstable}~\cite{WernerICRC21}. In particular, two major mechanical concepts are being considered for the construction of the core detector array: bladders installed in surface tanks, which could be made either of metal as in HAWC~\cite{HAWC17}, or rotomolded plastic as in the Pierre Auger Observatory~\cite{Allekotte2008}; and floating bladders directly deployed in a natural lake~\cite{GoksuICRC21} or an artificial pool. 

Regarding the common design elements, the requirement of an improved sub-TeV sensitivity implies the need to lower the energy threshold for detection of secondary shower particles (average energy $\sim$ 10\,MeV). For this, the use of reflective liners (tyvek internal layer) within the water volume enhances the number of detectable particles per air shower, improving trigger and energy reconstruction at lower energies. The presence of upward-facing PMTs are important for accurate timing within the individual unit, by measuring the direct Cherenkov light from particles entering the water volume, and inter-cell timing should be such as to preserve temporal accuracy. To achieve precision shower core reconstruction, not only the array fill factor, but also a compact detector unit size is important, as this increases the granularity of the array and the unit sensitive area, i.e. photon-sensitive area with respect to that of the detector unit footprint. The unit aspect ratio should also be optimised, constrained by the Cherenkov angle in water ($\sim 41^{\circ}$), with the requirement to maximise direct illumination of the PMTs at the base.

One possible design solution under study is the double-layered WCD ~\cite{Bisconti2022, Kunwar22}, with a $\gamma$-hadron separation strategy based on the use of vertical segmentation to identify energetic muons (typically few GeV) that penetrate the bottom layer. In addition, the bottom layer could provide a larger unit dynamic range, by enabling calorimetry even when the top layer PMT saturates in the case of energetic showers and high particle density. Another option under consideration is the use of shallow, multi-channel WCDs~\cite{MercedesWCD}. In this case, the design trade-off is between the reduced amount of water and the number of photo-sensors -- facing upwards at the bottom of the tank -- which would distinguish muons from electromagnetic particles based on the charge asymmetry between the PMTs, and rise times of the signals. Here, the asymmetry for muons is expected to be larger, as they traverse the entire water volume. Both options are in principle compatible with any of the mechanical implementations proposed. Adaptations may need to be considered in both designs for the outer array, in order to reduce scaling costs and avoid losses in the muon tagging efficiency.  

Concerning photosensors, large photo-active effective areas (of which vacuum PMTs provide the best relative cost) are needed to achieve single photo-electron threshold and lower the unit energy threshold, but electronics should be designed to avoid introducing bias in the shower reconstruction due to saturated detectors near the shower core. Nanosecond inter-unit timing accuracy and the ability to deal with high particle rates, especially at the highest altitudes, are other requirements to be taken into account for the readout and trigger electronics design.

Simulations are ongoing~\cite{Bisconti2022, Kunwar22, MercedesWCD} to evaluate the performance of the single WCD unit options with respect to size, aspect ratio, and PMT configurations, as well as the reflective material for the inner walls. Novel analysis strategies are also being considered~\cite{Conceicao21, Conceicao22}. The parameters taken into account to compare simulated unit performance are: number of photo-electrons produced, time resolution of the measurement of first photon, and detection efficiency, measured as fraction of particles entering the tank that produce a signal above threshold. Prototyping efforts are currently concentrated on the most critical items, aiming at collecting sufficient information to choose between the candidate options in Figure~\ref{fig:SWGOoptionstable}. Reliability of electronics and maintenance requirements at high-altitude sites is one such example. The robustness and easy deployment of the detector units is another issue under consideration, as well as bladder production, deployment and stability, which are especially critical for a lake-based solution.   

\begin{table}[t]
    \caption{South American candidate sites.}
\begin{tabular}{p{1.3cm} p{2cm} | p{1.3cm} p{1.3cm} p{4.5cm}} \toprule
    Country       & Site Name     & Latitude & Altitude   & Notes \\ 
                  &               &          & [m a.s.l.] & \\ \midrule\midrule
    Argentina     & Alto Tocomar  & 24.19 S  & 4,430      &  \\ 
                  & Cerro Vecar   & 24.19 S  & 4,800      & LLAMA, QUBIC \\ \midrule
    Bolivia       & Chacaltaya    & 16.23 S  & 4,740      & ALPACA \\ \midrule
    Chile         & Pajonales     & 22.57 S  & 4,600      & ALMA, Atacama Astronomical Park \\
                  & Pampa La Bola & 22.56 S  & 4,770      & ALMA, Atacama Astronomical Park \\ \midrule
    Peru          & Imata         & 15.50 S  & 4,450      & Represa Pillones\\ 
                  & Sibinacocha   & 13.51 S  & 4,900      & Lakes Cochachaca and Cocha Una \\
                  & Yanque        & 15.44S   & 4,800      & \\\bottomrule
\end{tabular}
\label{tab:SWGOsite}
\end{table}

\vspace{0.5cm}

As a final note, the Southern-Hemisphere location of SWGO will complement the reach of northern wide-field facilities such as LHAASO for an all-sky coverage, and will allow to fully exploit the synergies with CTA. In order to optmise overlap with LHAASO and to maximise the exposure to Galactic sources, and in particular the Galactic Centre ($\delta = -28.9^\circ$), SWGO is planned for installation in a latitude range between -15$^\circ$ and -30$^\circ$. That, in conjunction with the altitude constraints, for which a site above 4.4\,km a.s.l. is preferred, leave the South American Andes as the only viable option. Preliminary searches have found several candidates in Argentina, Bolivia, Chile and Peru (see Table~\ref{tab:SWGOsite}), each of which have specific strengths and match some array or detector technology options better than others. Overall, one of the main challenges for a final selection is water availability, of which an estimated $\sim$ 10$^5~\rm{m}^3$ will be required~\cite{DoroICRC21}.

\subsection{\textit{An Andean large-area particle detector for $\gamma$-rays - the ALPACA Experiment}}

The Andes Large area PArticle detector for Cosmic ray physics and Astronomy experiment, ALPACA~\cite{SakoICRC21}, will be devoted to the continuous observation and study of $\gamma$-ray signals from PeV cosmic-ray accelerators (PeVatrons) in the Galactic plane and the Galactic center. At a more advanced stage of development than other Southern Hemisphere proposals, it will be the first experiment to explore the Southern Hemisphere sub-PeV $\gamma$-ray sky. ALPACA will be a high duty-cycle, wide-field of view observatory sensitivity to $\gamma$-rays above several 10\,TeV, and capable of efficient particle identification and background rejection power in the UHE range thanks to an underground muon detector array.

The ALPACA Collaboration\footnote{\url{www.alpaca-experiment.org}} is an international project launched between Bolivia and Japan in 2016, and led by the Institute for Cosmic Ray Research (ICRR) of the University of Tokyo. It includes several member institutions from Mexico as well. The air shower array is currently being constructed at a high-altitude plateau, 4,740\,m a.s.l., near the Chacaltaya mountain in Bolivia, a historical site for cosmic-ray research in South America, on the outskirsts of La Paz. At such optimal altitudes the EAS of sub-PeV $\gamma$-rays have reached their maximum development. The plateau has a flat area of over 500\,m $\times$ 500\,m, ideal for installation of an extended array, as well as basic infrastructure (road, electricity) accessible nearby. The water source for the WCD units is currently under study, with a promising location identified within 1\,km from the site, where small lakes are present. Alternatively, underground water is available at 50 m depth. Truck transportation from the nearby town of El Alto is also possible, at a reasonable cost~\cite{SakoICRC21}. 

\begin{figure}[t]
    \centering
    \includegraphics[width=\textwidth]{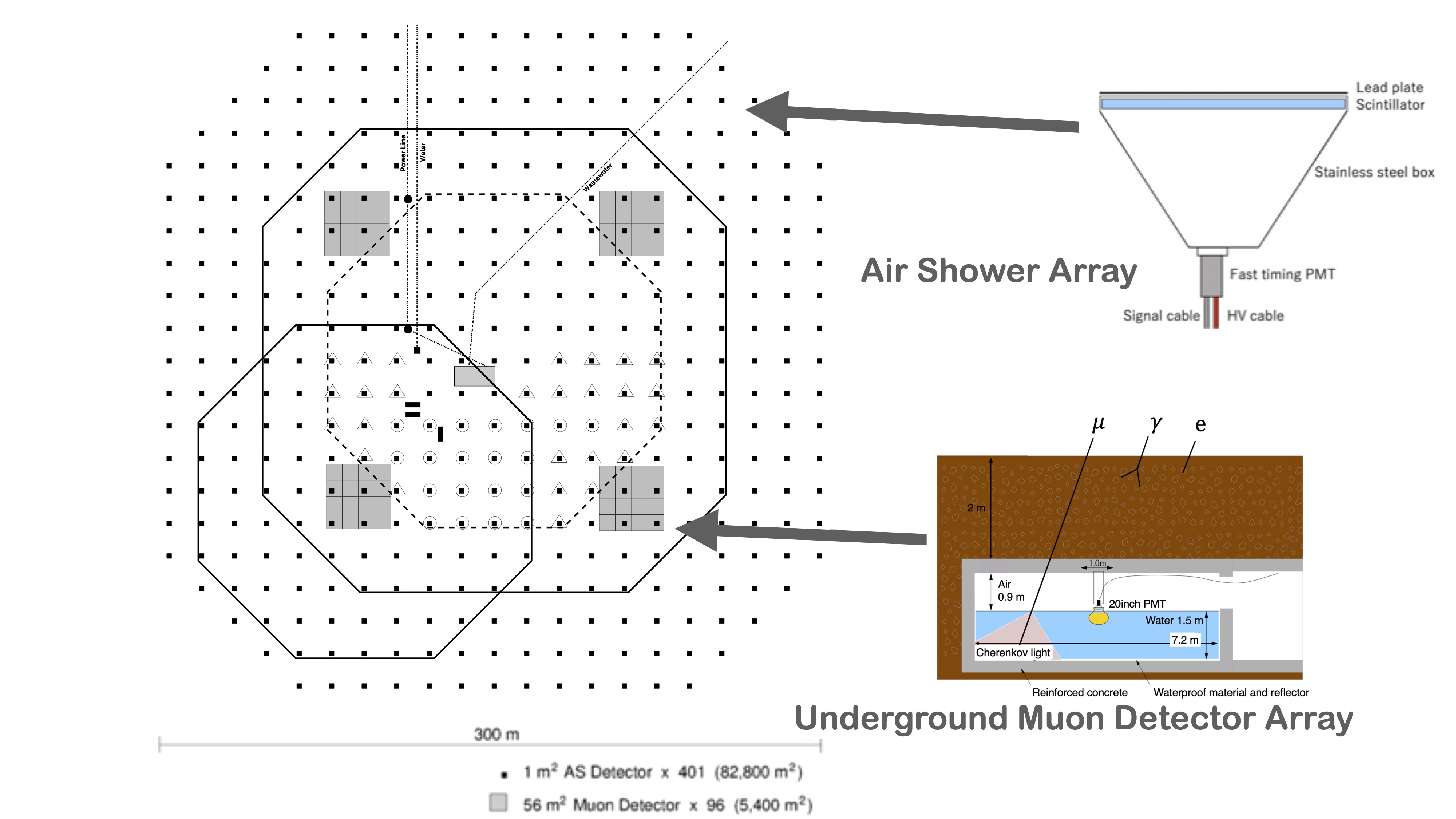}
    \caption{Layout of the full ALPACA array showing a depiction of the detector technologies used. Small dots indicate the 401 air shower detector units, space at 15\,m intervals. The enclosed area in the bottom left corner is the ALPAQUITA arrray. The half ALPACA has the same layout as ALPACA but with SD units spaced at 21\,m intervals. The muon detector consists of 4 clusters of 16 WCD units buried 2\,m underground. Image adapted from https://www.alpaca-experiment.org/as.html.}
    \label{fig:ALPACAarray}
\end{figure}

The ALPACA array (Figure~\ref{fig:ALPACAarray}) consists of two main components: the surface air shower detector array (SD), for energy reconstruction and timing of the shower front, and an underground muon detector array (MD). The presence of a dedicated muon array to select the muon-poor $\gamma$-induced showers greatly improves the sensitivity of the observatory, as validated by the similar concept applied in the Tibet-AS$\gamma$ experiment~\cite{Sako09}. Each SD unit comprises of 1\,m$^2$ active area $\times 5$\,cm thick plastic scintillator, with a 5\,mm thick lead plate on top, which is viewed by a fast-timing PMT. The lead is used to convert the EAS secondary $\gamma$-rays into pairs of $e^-$ and $e^+$ (Rossi transition effect) detectable by the scintillators. The MDs are water Cherenkov detectors placed 2\,m underground for an equivalent 16 radiation lengths shielding of the shower's electromagnetic component. An individual MD is a cluster of 16 cells, each with a 56\,m$^2$ area and 1.5\,m water depth which is viewed by a large 20" PMT placed on top. A muon $>$ 1\,GeV can penetrate the soil shielding and produce a clean, 24 photo-electrons signal on average, in the WCD PMTs. The pure muon signals allow for a 99.9\% hadron rejection, while retaining 80\% of the $\gamma$-ray signals above 100\,TeV. See Table~\ref{tab:ALPACAarray} for a detailed description of the array.

\begin{table}[t]
    \caption{Characteristics of the ALPACA experiment~\cite{SakoICRC21}.}
\begin{tabular}{p{4.cm} p{7cm}} \toprule
    Category                     & Description  \\ \midrule
    Location                     & Chacaltaya Plateau, Bolivia (68$^\circ$08'W, 16$^\circ$23'S)   \\
    Altitude                     & 4,740 m a.s.l. (572 g/cm$^2$)   \\
    Surface Area                 & 82,800 m$^2$ (fill factor 0.4\%) \\
    Muon Detector Area           & 5,400 m$^2$ (4 clusters, 16$\times$56 m$^2$ WCD cells) \\
    Surface Detector units       & 401 (1 m$^2$ $\times$ 5 cm, each)\\ 
    Energy resolution (0.1 PeV)  & 20\% \\
    Angular resolution (0.1 PeV) & 0.2$^\circ$ (50\% containment)\\\bottomrule
\end{tabular}
\label{tab:ALPACAarray}
\end{table}

Regarding the status of the project, the collaboration is now constructing a prototype array, ALPAQUITA, 1/4 of the full array area; it is expected to start operations in 2022, as soon as the first underground muon detector (MD) cluster is installed~\cite{Kato21}. Shortly afterwards, an extension to cover the full array area is scheduled (half ALPACA), with half the surface detector (SD) density and three additional underground MD clusters~\cite{YokoeICRC21}. The full ALPACA array is expected to be completed by the middle of the decade. The configurations of the array at each stage are given in Table~\ref{tab:ALPACAstages}.   

As show in Table~\ref{tab:ALPACAarray}, the expected full-array angular resolution (50\% containment) at 100 TeV is 0.2$^\circ$ (0.25$^\circ$ for half ALPACA~\cite{YokoeICRC21}), similar to that achieved by Tibet-AS$\gamma$~\cite{Amenomori21}. Simulations based on extrapolations of H.E.S.S. measurements of the Galactic Center~\cite{HESSGC} estimate that ALPACA should be able to detect $>$ 100\,TeV $\gamma$-rays from the GC after about 1-1.5 year of observations. Beyond ALPACA, the Collaboration plans a future extension towards a km$^2$ array (Mega ALPACA) to achieve sensitivity at the PeV energy range. The future extension should follow the same technology as currently applied.

\vspace{0.5cm}
{\bf ALPAQUITA}

The prototype array ALPAQUITA~\cite{Kato21} is currently under construction with the 97 SD units array completed by 2022. The surface array density of ALPAQUITA is the same as for the full array, with inter-unit spacing of 15\,m. Despite its reduced sized, of 18,450\,m$^2$, ALPAQUITA is not aimed as an engineering prototype only, with enough expected sensitivity to detect a few interesting sources. The array should become operational once construction of the first 900\,m$^2$ MD cluster is finalised, towards the end of 2022.

With a simulated energy resolution of $\sim$ 21\%, an angular resolution of $\sim 0.27^\circ$ (at 100\,TeV; 68\% containment), and an expected detection area of 12,600\,m$^2$ above 30\,TeV (showers at the edge of the array are not recorded) -- and thanks to the effective background rejection power of the underground muon array -- ALPAQUITA is expected to detect five Southern-Hemisphere sources seen by H.E.S.S., as an early validation for the first year of the experiment. In the case of the PeVatron candidate, HESS J1702-420A~\cite{GiuntiICRC21}, an unassociated hard-spectrum source seen by H.E.S.S. beyond 50\,TeV, ALPAQUITA should be able to extend spectral measurements beyond 300\,TeV, if no cut-off is present.

\begin{table}[t]
    \caption{Planning of ALPACA staging~\cite{SakoICRC21}.}
\begin{tabular}{p{2. cm} p{2.5cm} p{2.5cm} p{2.cm}} \toprule
    Stage         & Construction Year  & Surface Coverage    & MD clusters    \\ 
                  &                    & (Number of SDs)     & (1 = 16 cells) \\ \midrule
    ALPAQUITA     & 2022               & 18,450 m$^2$ (97)   & 1               \\ 
    half-ALPACA   & 2023               & 82,800 m$^2$ (200)  & 4               \\
    ALPACA        & 2024               & 82,800 m$^2$ (401)  & 4                \\
    Mega ALPACA   & after 2028         & 1 km$^2$ (1500)     & 50               \\ \bottomrule
\end{tabular}
\label{tab:ALPACAstages}
\end{table}

\subsection{\textit{The Cosmic Multiperspective Event Tracker (CoMET) Project}}

The Cosmic Multiperspective Event Tracker\footnote{\url{https://alto-gamma-ray-observatory.org}} (CoMET)~\cite{MezekICRC21} is a project for a wide field-of-view atmospheric shower array working in the VHE regime. The proposal is based on a hybrid design combining a $\sim 160$\,m diameter array of particle detector units (called ALTO) with atmospheric Cherenkov light collectors (CLiC stations), inspired by the HiSCORE design~\cite{T2014}. The array is to be placed at high-altitude, over 5 km a.s.l., to increase the capability to detect $\gamma$-rays down to 200 GeV. The main goal and key innovative aspect of the hybrid approach is to optimise the EAS sampling technique by improving shower reconstruction during the dark periods when the CLiC stations will be active, for better energy resolution and shower localisation. This is a key design element to achieve the proposed scientific objectives of CoMET, that is the study of soft-spectrum extragalactic $\gamma$-ray sources, such as Active Galactic Nuclei (AGN) and Gamma-ray Bursts (GRBs). The project is currently in R\&D Phase, and the prototypes of both the particle and atmospheric Cherenkov light detectors are under test at Linnaeus University, in Sweden. Concurrent simulations of the full detector response and shower reconstruction capability are underway.

\begin{figure}[t]
    \centering
    \includegraphics[width=\textwidth]{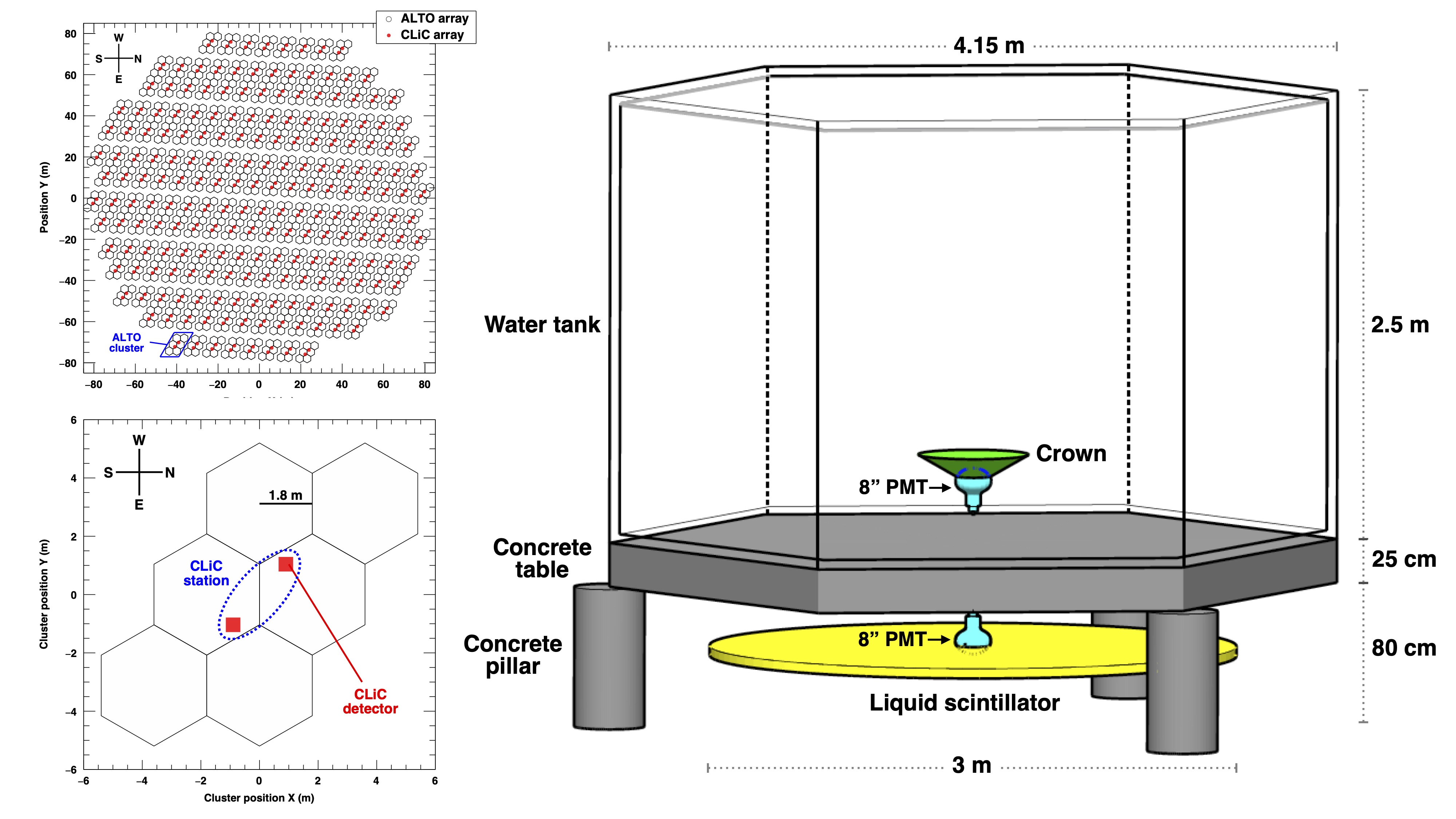}
    \caption{({\bf Top Left}) Layout of the CoMET array (ALTO and CLiC) showing ({\bf Bottom Left}) an overview of a single array cluster. ({\bf Right}) Depiction of an ALTO unit, consisting of a water Cherenkov detector hexagonal tank on top of a liquid scintillator for muon tagging. Image adapted from~\cite{MezekICRC21, SenniappanICRC21}.}
    \label{fig:CoMETarray}
\end{figure}

The experiment aims to install 1242 ALTO particle detector units, distributed in 207 clusters of 6 units over a circular area of $\sim 20,000~\rm{m}^2$. The particle array is complemented by 414 CLiC stations placed on top of the ALTO units, at a density of 2 atmospheric Cherenkov stations per particle array cluster (see Figure~\ref{fig:CoMETarray}). The two components of the experiment provide independent, but complementary information. The CLiC detectors are not aimed for array trigger, which will be based on the coincidence of WCD detectors. Shower core reconstruction is done by modelling of the lateral distribution function, as measured by the WCD stations. The timing from the WCD signals is used to model the shower front and reconstruct arrival direction. The reduced size of the tanks and closed-packing hexagonal design are important factors in achieving a fine sampling of the air-shower footprint at ground.

Shower development information from the CLiC stations can be used to further improve $\gamma$-ray source localisation during dark time. Simulations show that the addition of the CLiC results to the ALTO-only analysis results in a 10\% better angular resolution in the full energy range from 200 GeV to 100 TeV, and a 30\% better energy resolution around 1 TeV, with 12\% improved background suppression and negligible loss of $\gamma$-ray-like events. This is shown to have particular impact in improving detector sensitivity below 10 TeV~\cite{MezekICRC21}. The event analysis and $\gamma$-hadron separation strategies are described in~\cite{Senniappan21}. Below we give a quick outline of the individual CoMET detection units.

\vspace{0.3cm}
{\bf ALTO Stations}

Each ALTO unit~\cite{SenniappanICRC21} consists of a WCD, 2.5\,m high and 4.15\,m hexagonal-shaped tank, filled with $\sim 25~\rm{m}^3$ of water, and positioned over a 25\,cm concrete slab. Underneath this structure, a cylindrical liquid Scintillator Detector (SD), filled with Linear Alkyl Benzene (LAB), is placed (see Figure~\ref{fig:CoMETarray}). The WCD are primarily used for the detection of the secondary particles in the cascade, while the SDs are conceived for muon counting, with the concrete slab acting as an absorber to prevent the electromagnetic component from reaching the scintillator. The inner walls of the WCD are blackened to improve the timing accuracy from the PMT signal, thus helping to reconstruct shower direction. The use of closely-packed small tanks combined with more precise electronics and time-stamping are important proposed improvements with respect to current WCD designs, providing a fine-grained view of the shower particles at ground, for good arrival direction reconstruction and background rejection. The application of SDs also improve background rejection by muon-vetoing of cosmic-ray showers.

\vspace{0.3cm}
{\bf CLIC Stations}

The CLiC stations~\cite{MezekICRC21} are inspired in the HiSCORE wide-FoV technology used in the TAIGA experiment~\cite{Budnev2020,PorelliICRC21}, and consist of an array of 8 $\times$ 3" PMTs; this is different from the original HiSCORE setup which uses four channels consisting of large Winston cones coupled to 8" PMTs, and aims to improve the signal-to-noise ratio to low-energy atmospheric showers. Each CLiC 3" PMT is coupled to a 14\,cm Winston cone light guide, providing a final 0.1\,m$^2$ collection area per detector station, and an angular cut-off of $30^\circ$. A UV-pass filter is applied  to filter out the night sky background (NSB) and ambient light. The 16-channel signal from the two CLiC stations within a single ALTO cluster are finally summed into a single signal. 

Detailed simulations of event detection and analysis~\cite{Senniappan21} indicate an expected peak performance angular resolution of $\sim 0.15^\circ$ at 20\,TeV, which remains $< 1^\circ$ at energies as low as 300 GeV. Peak point flux sensitivity of $< 10^{-11} ~\rm{erg}.\rm{cm}^{-2}.\rm{s}^{-1}$ after 300 hours for near-zenith events is achieved around 5-10\,TeV. Towards the low-energies the expected sensitivity is $\sim 10^{-10} ~\rm{erg}.\rm{cm}^{-2}.\rm{s}^{-1}$ at 300\,GeV. This implies that the TeV-bright GRB 190114C could be detected within 30 min, a similar timescale than that needed to reach a 5-$\sigma$ detection of PKS 2155-304 in flaring state~\cite{SenniappanICRC21}. Further R\&D studies are underway, focusing particularly on improving the sensitivity $< 600$\,GeV by loosening the number of WCD stations needed for signal trigger and analysis. Prototyping efforts for achieving an improved timing accuracy through the CLiC stations is also underway.

\subsection{\textit{RPC-based proposals}}

\begin{figure}[t]
    \centering
    \includegraphics[width=\textwidth]{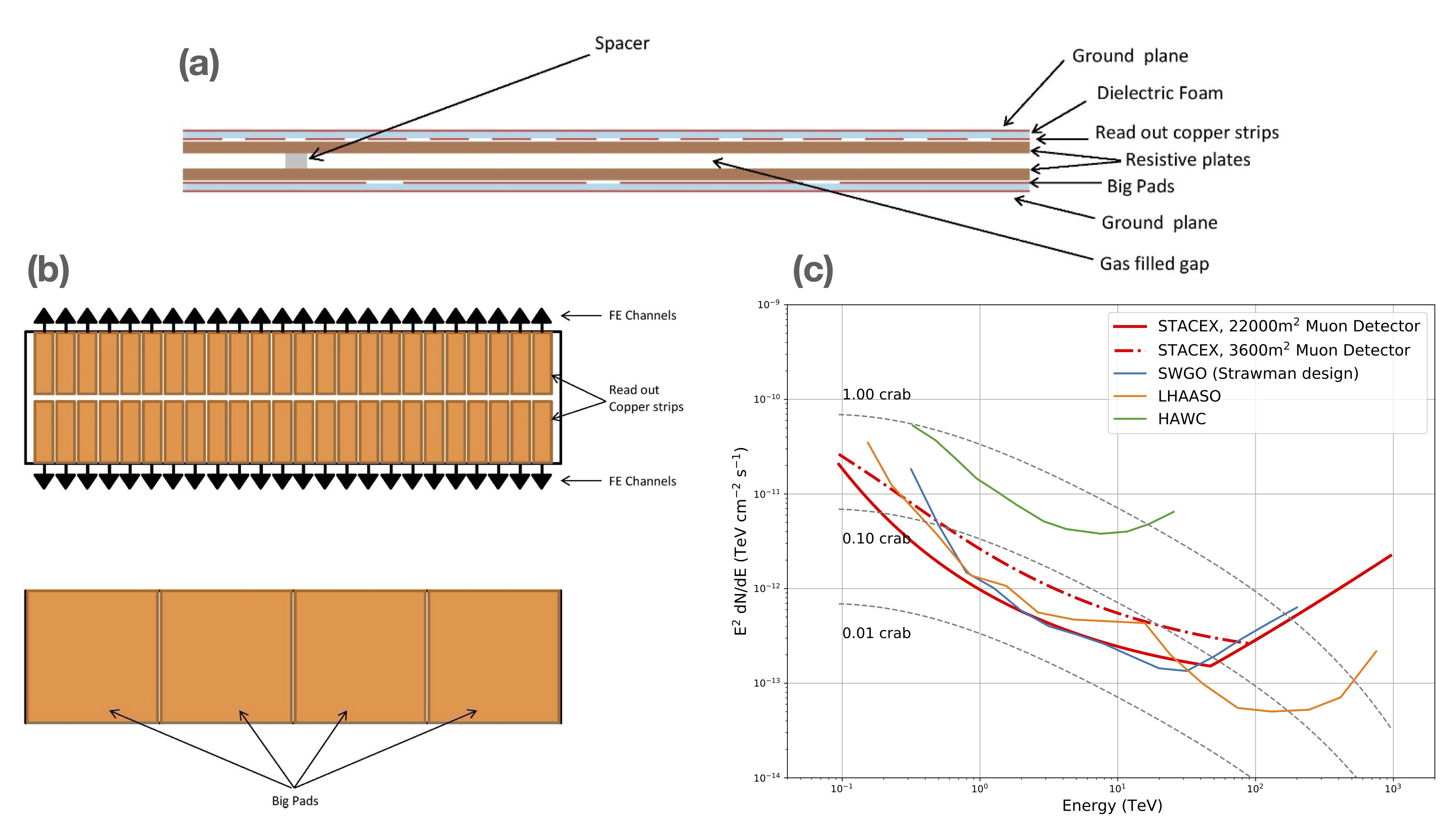}
    \caption{({\bf a}) RPC layout showing details of the detector elements. ({\bf b}) Structure of the read-out panels, as described in the text. ({\bf c}) Preliminary simulated sensitivity of STACEX for 1 year integration, compared to other experiments and proposed projects. Two layouts with different muon detector array areas are presented. Images taken from~\cite{DiSciascioICRC19, RodriguezICRC21}.}
    \label{fig:STACEX}
\end{figure}

Resistive Plate Chambers are among the types of detectors applied in particle sampling experiments, and have most notably provided the basis for the ground-based detector ARGO-YBJ, operational between 2007 and 2013 in Tibet, at an altitude of 4,300 m~\cite{DiSciascio14}. They are characterised by very high particle detection efficiency and timing accuracy. RPCs consist of a thin gas volume, where primary ionisation and avalanche multiplication occurs, sandwiched within a pair of resistive plates, where copper strips are placed to read out the signal generated inside the gas gap. The front-end electronics is embodied in the strip panels and the whole is sealed in mechanical support panels (see Figure~\ref{fig:STACEX}). In the case of ARGO-YBJ, the active elements consisted exclusively of RPCs, arranged as a full coverage (92\% fill factor) central carpet, 75$\times$75\,m$^2$ in area, and surrounded by a partially instrumented ring to improve the reconstruction of events with the core falling outside the carpet area. 

RPC-based detectors have some attractive features, which represent key instrumental strengths with respect to other approaches. From one side, the very dense sampling achievable with the RPC carpet can enable operation down to low energy thresholds, with very good position resolution, typically $\sim$ cm. Additionally, the high-granularity of the read-out of the RPC carpet can be exploited for very good energy and angular resolutions, and the flexible digital/charge read-out schemes opens the possibility for operation over a very wide dynamic range. 

Nevertheless, the intrinsic challenges of operating a gas-based detector at remote locations, and the associated costs of the technology, have traditionally limited its applicability and the achievable effective areas. In this sense, low gas consumption and simple electronics are desirable elements for an autonomous operation design that can balance the cost and power consumption constraints. Background discrimination capability, which is based on the study of the shower $>$ 40\,m away from the core, has also been limited by the small array sizes applied to this date in RPC-based experiments such as ARGO-YBJ, and consequently sensitivity has been historically underachieving. These factors are among the motivations for the use of RPCs in a hybrid approach.

The STACEX concept, which is currently the most relevant proposal to the continued development of RPC technology as a viable tool for ground-based $\gamma$-ray astronomy aims to improve the sensitivity of RPC-based arrays through such a hybrid approach using WCDs\footnote{An earlier proposal for a design combining RPC units and WCD tanks was the LATTES~\cite{LATTES18} project, which was nevertheless discontinued}.

\vspace{0.5cm}
{\bf The STACEX Concept}

STACEX~\cite{DiSciascioICRC19} is the concept for a high-altitude, hybrid detector composed of RPCs and water Cherenkov stations, which aims to achieve good shower reconstruction and $\gamma$/hadron separation over a very large dynamic range, from around 100\,GeV up to 10\,PeV. The team involved in developing the proposal is formed by scientists from the Italian National Institutes for Nuclear Physics (INFN) and for Astrophysics (INAF), responsible for previously operating ARGO-YBJ. The RPC detector proposed for STACEX is therefore similar to that used in ARGO-YBJ, but incorporates some evolution from the previous design, such as operation in avalanche, instead of streamer mode, as well as thinner electrode plates and new front-end electronics adequate to the avalanche mode operations.

The STACEX concept is based on a full-coverage ($\sim 90\%$) RPC carpet core, 150$\times$150\,m$^2$ in area, 4$\times$ that of ARGO-YBJ. An additional 0.5\,mm lead layer would be added on top of the RPCs to improve the quality of the temporal profile by exploiting the conversion of the secondary $\gamma$-rays. The arrangement is complemented by a dense muon detector array placed below the carpet, constituted by WCD tanks buried under 2.5\,m of soil, akin to the muon-array of LHAASO-KM2A. The performance target aims at a very good energy ($<$ 20\%) and angular (better than 0.2$^\circ$) resolutions above 10\,TeV. Further expansion of the dynamic range beyond 100\,TeV could be achieved by the addition of a few \% active area scintillator-based outer array, covering a total area of 0.5\,km$ ^2$ around the core. Together with the capability of detecting clear muon signals with buried WCDs, the concept could guarantee a high-efficiency rejection of cosmic-rays and multi-PeV sensitivity.

As currently proposed, the carpet would follow the same design used in ARGO-YBJ, of an array of bakelite-based RPCs. The proposed detector has a modular structure, with clusters of 5.7$\times$7.6\,m$^2$, made of 12 RPCs of 2.85$\times$1.23\,m$^2$ each. The high-granularity read out is made by 80 external strips per chamber (for a total of 570,216 strips), defining a spatial pixel of 6.75$\times$61.80\,cm$^2$, logically organised in 10 independent pads of 55.6$\times$61.8\,cm$^2$ (for a total of 71,277 pads), which are the temporal pixels of the detector. In order to extend the dynamic range to the PeV, each chamber is equipped with two large size pads (139$\times$123\,cm$^2$) to collect the total charge of the particles hitting the detector~\cite{RodriguezICRC21}. 

Figure~\ref{fig:STACEX} illustrates the structure of the RPC read-out panels, and shows the expected sensitivity of a STACEX-like array of 22,000\,m$^2$ in area. The event selection criteria for the analysis required minimum trigger of 20 strips on the carpet and a reconstructed core position inside an area of 600$\times$600\,m$^2$ centered on the detector. Core reconstruction by the RPC carpet has a resolution of 20\,m at 100\,GeV, down to 2\,m at 100\,TeV. The lowest energy bin reconstructed for a strip multiplicity between 20 to 40 strips, is $\sim$ 100\,GeV, with 50\% resolution~\cite{RodriguezICRC21}.

Background rejection in the STACEX concept relies on the underground WCD array to reject charged CR showers based on their muon content, and the efficacy of the approach depends both on the size and coverage of the muon array and the muon detector efficiency. Figure~\ref{fig:STACEX} presents the sensitivity for two distinct muon array layouts, with continuous (22,000\,m$^2$) and partial (3,600\,m$^2$) core coverage. It is clear that the size of the muon detector has a significant impact in sensitivity, and that a large muon array is needed to reduce the sampling fluctuations of hadron-initiated showers. A large overall detector is also important to exploit the pattern of energy deposition in the surface detector, away from the shower core ($\gtrsim$ 40\,m), for rejection of hadron-initiated showers. Preliminary studies suggest that the hybrid approach could reach the background-free regime, with a background rejection level of 3$\times$10$^{-4}$ above few 10s of TeV, and nearly 100\% $\gamma$ survival rate, thanks to the very dense sampling achieved by the dense core (RPC carpet plus muon array). As already mentioned, such performance could be expanded beyond 100 TeV provided that a suitable extension of the array is available to reach the required photon statistics~\cite{RodriguezICRC21}.

Up until 2021, the STACEX proposal had conducted only preliminary simulation studies, which nevertheless suggested that RPCs could be a suitable technology for a wide-field $\gamma$-ray detector in the Southern Hemisphere, provided that a sensitive muon array is operated in conjunction.

\section{Future Imaging Atmospheric Cherenkov Experiments}

The experimental context of ground-based gamma-ray astronomy today is dominated by the imminent start of construction of the Cherenkov Telescope Array (CTA), complemented, at the UHE regime, by the significant advances brought by LHAASO from the side of the EAS arrays. One of the principal technical advantages of IACTs over the particle samplers discussed in previous sections are their excellent $\gamma$/hadron discrimination capability, resulting from the Cherenkov-light imaging of the complete shower development in the atmosphere. Another is the very large effective area, even for a single telescope, which corresponds to the size of the Cherenkov light pool produced by the showers at ground, $\sim 10^5~\rm{m}^2$. That said, above 100\,TeV, the imaging atmospheric Cherenkov technique loses part of its competitiveness, and the shower sampling arrays stand out as the most viable approach, being able to cover the $\sim\rm{km}^2$ ground areas and steradian angular fields required at a lower cost, while operating with a much more favourable duty cycle.

Looking into the future of the air-Cherenkov experiments, a few crucial instrumental frontiers have been identified: 

\begin{itemize}
\item larger arrays for improved sensitivity at multi-TeV energies and short timescales; 
\item enhanced angular resolution for morphological studies of extended sources and improved isotropic cosmic-ray background suppression;
\item lower energy threshold, towards few tens of GeV, for extra-galactic and variable / multi-messenger science;
\item wider field of view, for the conduction of surveys, study of extended sources and diffuse emission, as well as the search for transient / serendipitous phenomena
\end{itemize}

All these goals will be addressed by CTA, the next-generation ground-based gamma-ray observatory, which will deliver an order of magnitude improvement in all performance parameters with respect to current instruments. But even in the context of CTA, other IACT experiments are being proposed, focusing on the development of particular instrumental frontiers and aiming to provide scientific contributions at specific areas.

\subsection{\textit{The CTA Context}}

An ideal IACT array uses many telescopes to densely sample the air-Cherenkov light pool, determining the shower properties with great precision, and enhancing the capability to separate the hadronic background to achieve excellent sensitivity. With a planned full-array of over 100 IACTs, deployed over a multi-km$^2$ area in two sites on both hemispheres, CTA will be the definitive observatory for VHE astronomy in the coming decades, reaching a point-source sensitivity of 0.1\% of the Crab Nebula within 50 hours integration time. Furthermore, by combining different sizes of telescopes and covering a large ground area, CTA will expand the performance of current facilities over a very broad range, 4 decades in energy. CTA will distinguish itself in the network of ground-based $\gamma$-ray instruments by partly functioning as an open observatory to the scientific community, in a field where facilities were traditionally run as closed experiments. For more on the Cherenkov Telescope Array, see chapter "The Cherenkov Telescope Array (CTA): a worldwide endeavour for the next level of ground-based gamma-ray astronomy".

In the context of the technological R\&D for CTA, a number of telescopes and camera prototypes have been developed by various groups over the past decade. Among them, the ASTRI design by INAF (the Italian National Astrophysical Institute) advanced a pioneering Schwarzchild-Couder configuration which, beyond working as a prototype for the CTA small-sized telescope, will compose an independent "pathfinder" Mini-Array, whose operation will precede that of the full CTA array. Scientifically, it will complement current facilities by significantly extending the energy reach of Northern Hemisphere IACTs before CTA is fully operational. It will also allow the early exploration of important synergies with LHAASO over an overlapping band of several tens of TeV. The case of the ASTRI Mini-Array demonstrates that there still exists a meaningful role to be played by novel IACT facilities in the era of CTA, which can operate as experiments focused on specific science goals, and complement the capabilities of CTA in terms of temporal, geographical, and spatial coverage. They can also expand some of CTA's technological frontiers, and serve as test beds for future developments in the field.

Another relevant case is MACE, the recently installed 21\,m diameter Indian IACT. MACE is a state-of-the-art instrument and the latest step in the long development of ground-based $\gamma$-ray astronomy in India. Its most distinguishing features are its geographical location (the easternmost IACT in the world) and high installation site -- at 4,270 m a.s.l., it is the highest IACT ever built. The idea of installing a high-altitude IACT aims at pushing the observational threshold to the lowest possible energies, down to a few tens of GeV. 

As the density of Cherenkov light from the air-shower increases monotonically with elevation, the installation of an IACT at very-high altitude allows the detection energy threshold to be directly reduced of the detection energy threshold. But the observation of low-energy showers introduces experimental challenges of its own. From one side, very-low energy showers of $\sim 10$ GeV emit Cherenkov radiation mostly in the first generations of secondary electrons (which are above the $\sim$ MeV threshold for Cherenkov light production) and not throughout the cascade development, as is the case for higher-energy primaries. This means that the Cherenkov image is less regular and more susceptible to fluctuations, and has an important impact in increasing both the PSF and the energy bias and resolution at the lowest energies. The high altitudes also affect the reconstruction capabilities for higher energy showers, since the shower must die completely before reaching the ground for the Cherenkov emission to provide an effective calorimetric measurement. On the other hand, the high altitude results in a non-negligible reduction of the amount of light produced from hadronic showers, that penetrate deeper in the atmosphere, favouring background rejection. 

Experimentally, the energy range from 10 to 100 GeV constitutes a challenging and traditionally less explored spectral region, bridging between the satellite and ground-based observational regimes, and approaching the few-GeV theoretical limit for air-Cherenkov observations. Thanks to the combination of higher source fluxes and the large collection areas at ground, the science potential at this energy regime is strongly focused on variable phenomena, the instrument serving as an ideal "trasient explorer".

\subsection{\textit{ASTRI}}

\begin{figure}[t]
    \centering
    \includegraphics[width=\textwidth]{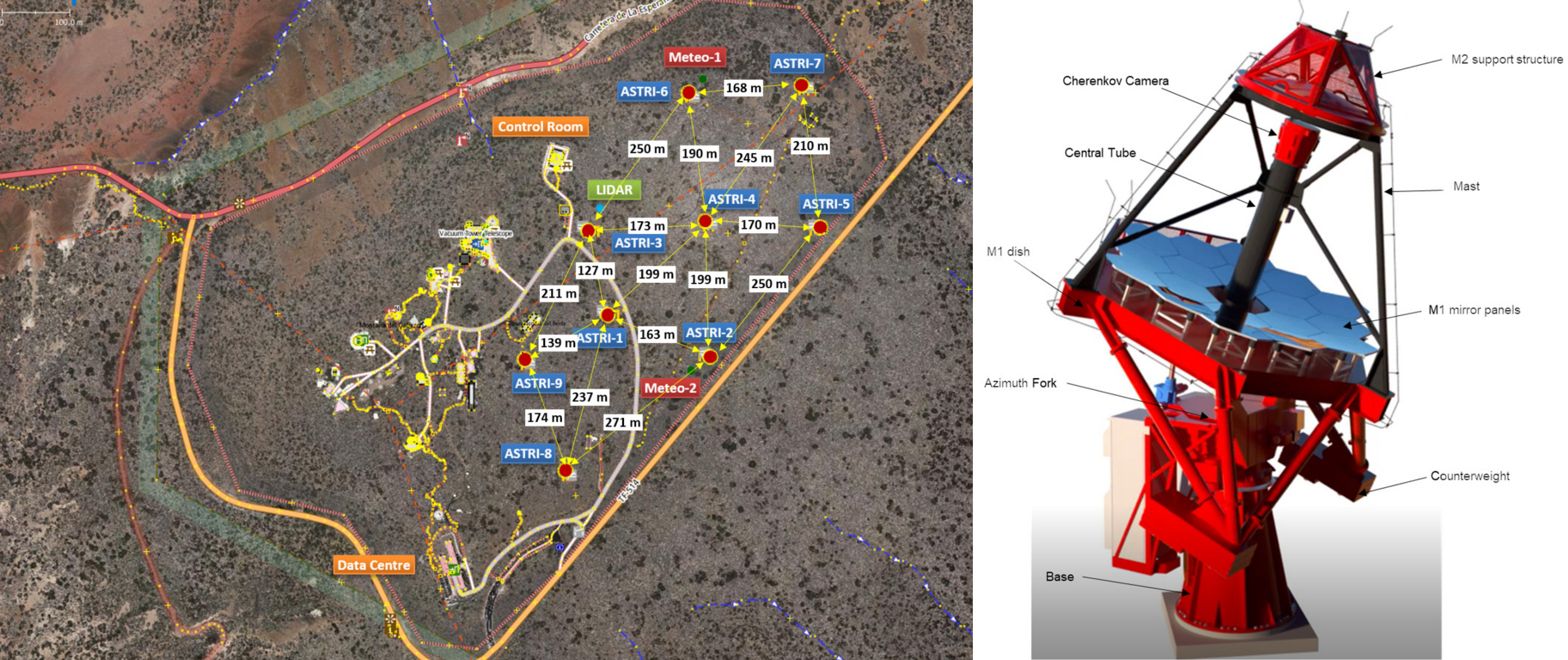}
    \caption{({\bf Left}) A schematic view of the final array layout for the 9-telescope ASTRI Mini-Array at Teide Observatory. ({\bf Right}) The model of the configuration of a telescope of the ASTRI Mini-Array. Images taken from~\cite{AntonelliICRC21}.}
    \label{fig:ASTRI}
\end{figure}

The ASTRI project is a future observatory to be installed at the Teide Observatory, in Tenerife, dedicated to the study of the VHE $\gamma$-ray sky in the range from a few TeV to 100 TeV and beyond. The innovative telescope technology is based on the ASTRI-Horn design~\cite{Lombardi20}, developed by INAF within the context of CTA, as a prototype 4-m diameter small-sized telescope (SST) (see Figure~\ref{fig:ASTRI}). The ASTRI-Horn\footnote{In honour of the Italian-Jewish astronomer Guido Horn D'Arturo, who pioneered the use of segmented primary mirror in astronomy~\cite{Horn36}.} prototype is currently installed and operational at the INAF observing station at Mt. Etna, in Sicily. It is the first dual-mirror optical configuration Cherenkov telescope ever deployed, and among the first to use Silicon Photo-Multipliers (SiPM) as detectors. The ASTRI prototype has achieved first light in December 2018, with the detection of the Crab Nebula above 3.5 TeV~\cite{Lombardi20}.

The optical design is the main technological innovation of the ASTRI-Horn telescope. Typically, IACT experiments use tessellated, single mirror systems based either on the Davies-Cotton~\cite{Davies57} or parabolic~\cite{Bastieri05} configurations. Although convenient for IACT Astronomy, where isochronicity of the signal is important, single mirror designs have a limited field of view and significant off-axis aberration, and also imply the use of bulky cameras placed at the focal plane, due to the resulting large plate-scale. Dual-mirror configurations, such as the Schwarzchild-Couder (SC) aplanatic design proposed by~\cite{Vassiliev07} and pioneered in the ASTRI-Horn telescope, allow to implement a large field of view, up to $10^\circ$ in diameter, in a more compact instrument, with consequently much smaller plate-scale, thus preserving a good angular resolution throughout the entire FoV. As a result, the dual-mirror design enables for the correction of aberrations at large-field angles.

The primary mirror of the ASTRI-Horn telescope has a diameter of 4.3\,m, composed of 18 hexagonal tiles, while the secondary reflector consists of a 1.8\,m diameter monolithic hemispherical mirror with 2.2\,m radius of curvature. The configuration results in a focal length of 2.15\,m (f/0.5) and allows to cover a full FoV larger than $10^\circ$. The camera plate scale is small, 37.5\,mm/$^\circ$. Such optical properties represent significant steps ahead in the design of Cherenkov telescopes, resulting in compact cameras and the possibility to use small SiPM pixel sizes (7\,mm linear dimension in this case) as an alternative to the traditional PMTs. The SiPM cameras of the ASTRI-Horn telescope also follow a novel, curved focal-plane camera design, with very-fast read-out electronics~\cite{Catalano18}. The application of SiPM cameras in IACTs has already been successfully demonstrated by the FACT telescope~\cite{Anderhub13}.

The science commissioning phase of the ASTRI-Horn telescope has started in 2020, and is ongoing, with hardware improvements on the optical system and camera scheduled to reach nominal configuration. The goal is to move towards the second phase of the ASTRI project, represented by the development of a Mini-Array composed of nine dual-mirror IACTs, to be installed at Teidei Observatory, in Tenerife, Canary Islands.

\vspace{0.5cm}
{\bf The ASTRI Mini-array}

The goal of the ASTRI Mini-Array~\cite{AntonelliICRC21}, to be jointly operated by INAF and the Instituto de Astrofisica de Canarias (IAC), will be to carry out observations of $\gamma$-ray sources up to 200\,TeV, extending significantly the energy range explored by current IACTs. It aims to exploit the large FoV of the dual-mirror telescope design to discover serendipitous transient sources, simultaneously perform deep exposures of a number of selected targets and fields, as well as to study large extended objects such as Supernova Remnants (SNR) and Pulsar-Wind Nebulae (PWN). In fact, the excellent angular resolution of the observatory, between 0.05$^\circ$-0.15$^\circ$, means it will be able to study in unprecedented detail the morphology of such sources above 10\,TeV. Given the relatively high energy threshold of the observatory, of 2\,TeV, and the fact that it will operate in a context defined by the start of operations of CTA, the observational schedule of the ASTRI Mini-Array will focus on deep exposures of selected, science-driven targets~\cite{Pintore20}. Many relevant synergies with the EAS arrays HAWC and LHAASO, continuously surveying the northern sky in the VHE-UHE regimes, are also expected. 

Initially conceived as a pathfinder array for the CTA-South at Paranal, in Chile, the ASTRI Mini-Array is now an independent project, with a specific set of science goals, and expected to begin operations in 2024. Once operational, it will be the most sensitive among current IACT arrays above $\sim$ 10\,TeV, thus complementing CTA in performing unprecedented detailed observations of the $\sim$ 10-200 TeV $\gamma$-ray sky. As such, the ASTRI Mini-Array will devote the first years of its observing programme to deep and ultra-deep observations (200-500 hour exposures) of core science topics, aiming to address a few relevant open questions, such as~\cite{VercelloneICRC21}:   

\begin{itemize}
\item testing the "hadron-beam" scenario of blazar emission, and the connection between blazar jets and UHECRs;
\item probing the EBL IR component through observations above 10 TeV;
\item performing deep probes of the Galactic Center at high zenith angles, exploring its excellent angular resolution to identify PeV particle accelerators;
\item performing good angular resolution observations of LHAASO-discovered PeVatron signals, to study their morphology below 100\,TeV and attempt an unequivocal identification of their astrophysical counterparts.
\end{itemize}

\begin{figure}[t]
    \centering
    \includegraphics[width=\textwidth]{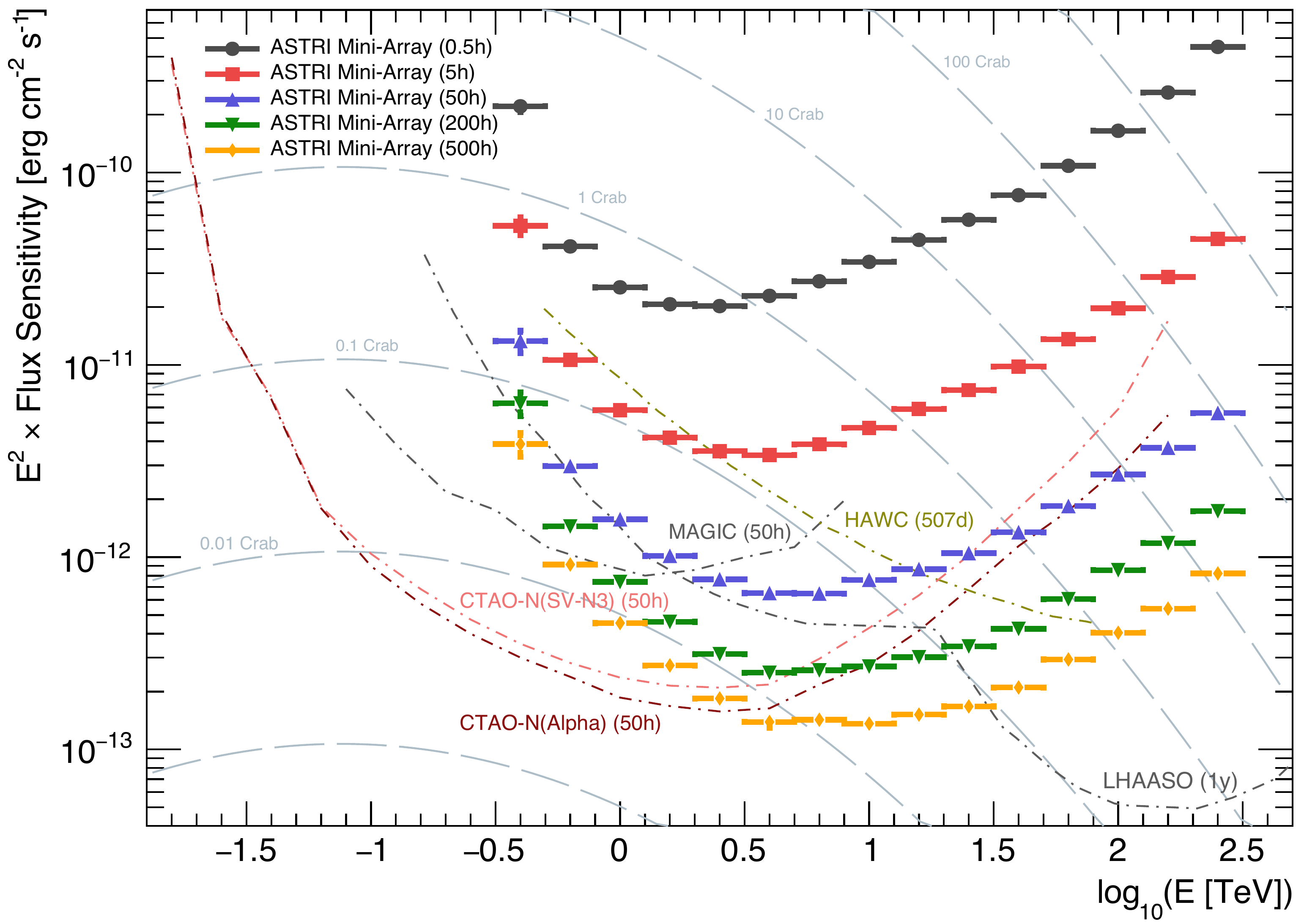}
    \caption{Point-source differential sensitivity of the ASTRI Mini-Array (calculated at 20$^\circ$ zenith angle) for different exposure times, from 30\,min snapshots to the deep and ultra-deep exposures (200 and 500 hours) planned for the core science topics. In comparison are shown the sensitivity curves for other northern-hemisphere ground-based instruments: MAGIC (50 hours), HAWC (507 days), LHAASO (1 year) and CTA-N (50 hours). The Crab Nebula model is also shown, scaled to different flux levels. Images taken from~\cite{LombardiICRC21}.}
    \label{fig:Mini-Array}
\end{figure}

The final layout of the 9-telescope Mini-Array is shown in Figure~\ref{fig:ASTRI}. It will consist of an NE-SW elongated asymmetrical layout with a median spacing among telescopes of $\sim$ 200\,m, near-optimal for operations $>$ 10\,TeV. Figure~\ref{fig:Mini-Array} shows the expected differential sensitivity curve of the ASTRI Mini-Array~\cite{LombardiICRC21}. It can be seen that for the deep (200\,h) and ultra-deep (500\,h) exposures of the core science programme~\cite{VercelloneICRC21}, the Mini-Array will achieve exceptional sensitivities above 10\,TeV (higher than CTA for 50\,h exposures), and will in any case be more sensitive than current IACTs above a few TeV, for typical integration times of 50 hours. The on-axis angular resolution (68\% containment radius) will be as good as 0.05$^\circ$ in the range from a few TeV to 100\,TeV, with an energy resolution of the order of 10\% for this same range. The ASTRI Mini-Array will thus be a significant complement to both present-day and next-generation observatories in the Northern Hemisphere.

\subsection{\textit{MACE}}

MACE (Major Atmospheric Cherenkov Experiment) is the future Indian telescope dedicated to ground-based $\gamma$-ray astronomy. It represents a continuation in the regional development of the field after two decades of activities of its predecessor instrument, the TACTIC array ~\cite{Singh21}. The telescope has been recently installed, and is under commissioning~\cite{YadavICRC21} at the Himalayan Gamma ray Observatory (HIGRO), in Hanle, Northern India ($32.8^{\circ}$ N, $78.9^{\circ}$ E). Its geographical location fills an important longitudinal gap among current and future IACT instruments in the Northern Hemisphere. At a site 4,270\,m in altitude, MACE will be the highest existing IACT in the world. Although MACE is a single-telescope facility, it closely follows an early proposal for a high-altitude IACT observatory~\cite{Aharonian01} and puts forward the case for the installation of a low-energy stereoscopic array at high altitude.

\begin{figure}[t]
    \centering
    \includegraphics[width=\textwidth]{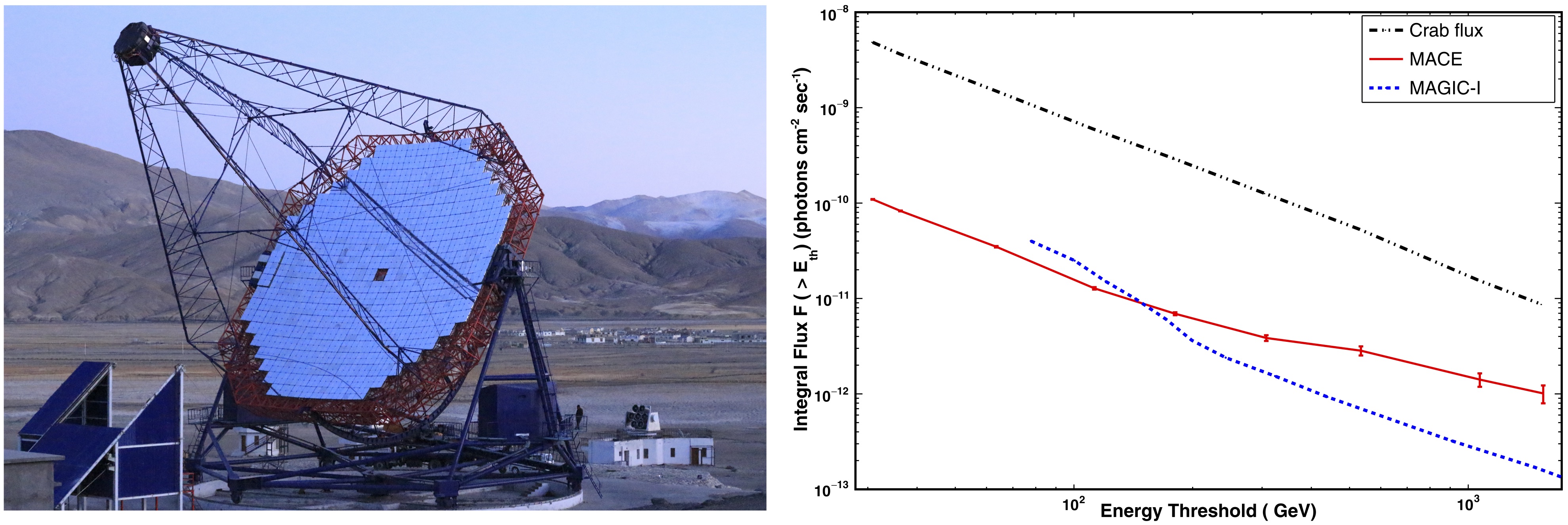}
    \caption{({\bf Left}) Image of the 21 m diameter MACE $\gamma$-ray telescope at Hanle site. ({\bf Right}) Integral flux sensitivity of the MACE telescope at zenith $< 30^\circ$, with MAGIC-I sensitivity shown for comparison. From~\cite{Singh21}.}
    \label{fig:MACE}
\end{figure}

An image of the MACE telescope is shown in Figure~\ref{fig:MACE}. The instrument has an alt-azimuth mount, with a quasi-parabolic light collecting surface of 21\,m in diameter, comprised of 356 independent mirror panels of 1\,m $\times$ 1\,m each, and 25\,m focal length. Such a large optical reflector is a basic design element, allowing to collect as much Cherenkov light as possible from the weak low-energy showers and effectively lower the observational energy threshold of the instrument.  Each panel is in turn composed by four spherical aluminum honeycomb mirror facets of 0.5\,m $\times$ 0.5\,m (in a total of 1424), with graded individual focal lengths between 25\,m and 26.5\,m, from the centre to the periphery of the reflective surface, in order to guarantee a minimum on-axis spot size at the focal length of the telescope. The reflectance of the mirror facets is superior to 85\% in the wavelength range of interest for Cherenkov light, between 280-700 nm. The mirror facets are aligned by an active mirror control system. The total mirror collection area of the telescope is $\sim 346$\,m$^{2}$. The quasi-parabolic design is chosen to reduce the optical aberrations of the large reflector surface.

 For such a high-altitude instrument, the camera can be more compact, and achieve the necessary angular resolution for imaging with a relatively smaller number of pixels. The MACE telescope camera is placed at the focal length of the reflective surface and has a modular structure, with 68 modules with in-built digital signal processing electronics, each consisting of 16 photo-multiplier tubes (pixels) of 38\,mm in diameter. All of the 1088 PMTs are fitted with hexagonal parabolic light guides, each with an angular size of 0.125$^\circ$, for an effective coverage of the entire surface area of the camera, and a total field of view of $\sim 4.3^\circ \times 4.0^\circ$. Along with the large mirror area, such a multi-channel, high resolution camera, is essential for performing a good imaging of the higher granularity\footnote{Defined as the average angular separation between points of emission of Cherenkov light during the development of extensive air-showes} low-energy Cherenkov images at high-altitude, as needed for a good $\gamma$-hadron separation. Only the 36 inner modules (576 pixels) will be used for event trigger, with a field of view $2.6^\circ \times 3.0^\circ$. A trigger configuration of 4 close-cluster nearest-neighbour pixels is implemented in the MACE hardware. For a large telescope aperture as in MACE, such a trigger strategy is relevant in order to effectively suppress accidental events from the night-sky background (NSB) and thus allow the instrument to fully explore the increase in the number of photo-electrons registered per shower that results from the high-altitude installation.

As pointed out earlier, the very high altitude site of MACE brings along two important advantages to be explored in reducing the observational threshold and improving low-energy sensitivity, as it both reduces the absorption of the shower Cherenkov light by the atmosphere, and increases the density of the Cherenkov photons at ground (to $\sim 1$\,ph/cm$^{2}$ for a 10\,GeV $\gamma$-ray shower). This geometrical increase in the Cherenkov photon density due to the altitude is also more pronounced for $\gamma$-ray than hadronic showers, which are more penetrating, and weights favourably in the trigger probability of signal over background events, especially below 100\,GeV. As a result, MACE can achieve a very low $\gamma$-ray trigger threshold of about 20\,GeV in the low-zenith angle range $< 40^\circ$, leading to an excellent single-telescope integral sensitivity for energies below $\sim 150$\,GeV~\cite{Borwankar20}. The 50\,h integral sensitivity of the telescope above the analysis threshold of $\sim$ 30\,GeV is estimated in 2.7\% of the Crab~\cite{Sharma17} (see Figure~\ref{fig:MACE}). The expected energy resolution of the instrument is about 40\% below $\sim$ 50\,GeV, improving to $\sim$ 20\% around 1\,TeV; the angular resolution similarly improves from $\sim 0.21^\circ$ below 50\,GeV, to $\sim 0.06^\circ$ at few TeV. The dynamic range of operations of MACE will be between $\sim$ 30\,GeV - 10\,TeV~\cite{Borwankar20}.

Owing to its high altitude site and large collection area, MACE will be focused on the study of sources in the 20-100\,GeV energy region, mostly unexplored by ground-based instruments, with a science case strongly centred around the observation of bright flaring objects such as AGNs. Here, thanks to its eastern location, it can valuably complement CTA by extending the monitoring coverage capabilities. The very-low energy threshold of MACE, as well as its peak differential sensitivity around 100\,GeV, will provide it with a unique overlapping range with satellite observatories such as Fermi-LAT. This should allow to fill some gaps with respect to other IACTs in the spectral and temporal studies of certain classes of sources $< 100$\,GeV, such as in the observation of distant extragalactic sources for EBL measurements, and on the observation of the GeV-TeV component of pulsars, profiting of the much larger collection area of ground-based instruments in comparison to satellites.

Since the conclusion of its installation at the end of 2020, MACE has been under commissioning, performing trial runs for testing the performance of different telescope components~\cite{YadavICRC21}, and first light is expected soon.

\section{Outlook}

The current (third) generation of IACT instruments (HESS, MAGIC and VERITAS) has brought ground-based $\gamma$-ray astronomy to maturity, improving sensitivity in the TeV range by over an order of magnitude with respect to previous instruments, while lowering the observational threshold to under 100\,GeV. As a result, the number of known TeV sources has dramatically increased in the past 20 years. Among the technical ingredients behind these achievements were the perfecting of the stereoscopic imaging technique, and the construction of large reflectors, with over 100\,m$^2$ in area, as well as cameras with a large FoV and fine-pixel sizes $\lesssim 0.2^\circ$, which permitted a good resolution of shower features. The imminent coming online of CTA will represent an even larger step in evolution and the inauguration of a new era of astrophysical research for this now fully-established field of astronomy.

The past decade has also seen the flourishing of ground-breaking particle arrays such as HAWC and, more recently LHAASO, which together detected over 50 sources above several tens of TeV, finally opening up the UHE astronomical window. In this case, the installation of large arrays (much larger than the shower footprint) at high altitudes, above 4 km a.s.l., with dense instrumented areas that permitted a good calorimetric measurement of the showers and timing of the shower front, were crucial, along with the large muon effective detection areas, which enabled an excellent $\gamma$-hadron separation.

As described in this chapter, the next steps will centre around hybrid arrays, as in the case of TAIGA, which is focusing on cost-effective ways to instrument very large areas while retaining at the same time a good reconstruction quality and $\gamma$-hadron separation power. Another frontier is the expansion of the wide-field coverage towards the Southern Hemisphere, a still unexplored region of the sky for ground particle arrays. Here, SWGO figures as the principal proposal, bringing together all crucial elements such as a very large array area for the consolidation of the (sub-)PeV observational window, the expansion of the energy reach towards lower energies, and the improvement of shower reconstruction, aiming particularly at a better angular resolution. 

The plans for new instrumentation are at various stages of maturity, but the emerging scene for the field within this decade is that of a well-established global network of ground-based $\gamma$-ray instruments, with the potential to work in close synergy to extract the most from the various complementary observational techniques available. Together, these instruments will not only be at the forefront of high-energy astrophysics and astro-particle physics, but will be an essential ingredient of the nascent multi-messenger astronomy. 

In summary, the field of ground-based $\gamma$-ray astronomy is very different from even a decade ago, where discussions revolved around expectations on the number and classes of sources available for detection by the instruments of that time. Today, the scientific implications of the field have clearly taken centre-stage and are the drivers of the next steps, and we can expect an even greater progress for the field within the next ten years. Overall, the instrumental prospects have never been richer, with concepts for new facilities being presented throughout the energy domain and across the globe, and evoke a well-known quote by Pierre Theillard de Chardin: "The history of science can be summarised as the development of ever more perfect eyes in a world where there is always more to see."

\newpage

\section{Cross-References}

\begin{itemize}
\item How to detect Gamma-rays from ground: an introduction to the detection concepts
\item The development of ground-based Gamma-ray astronomy: a historical overview of the pioneering experiments
\item Detecting gamma-rays with high resolution and moderate field of view: the air Cherenkov technique
\item Detecting gamma-rays with moderate resolution and large field of view: Particle detector arrays and water Cherenkov technique
\item Current particle detector arrays in gamma-ray astronomy
\item The Cherenkov Telescope array (CTA): a worldwide endeavor for the next level of ground- based gamma-ray astronomy
\end{itemize}

\end{document}